\documentclass[twocolumn]{aastex631}

\usepackage{float}

\begin{document}

\title{ALMA High-resolution Spectral Survey of Thioformaldehyde (H$_{2}$CS) Towards Massive Protoclusters}

\author[0009-0009-8154-4205]{Li Chen}
\affiliation{School of Physics and Astronomy, Yunnan University, Kunming 650091, People’s Republic of China}
\correspondingauthor{Li Chen}
\email{li.chen@mail.ynu.edu.cn}

\author[0000-0003-2302-0613]{Sheng-Li Qin}
\affiliation{School of Physics and Astronomy, Yunnan University, Kunming 650091, People’s Republic of China}

\author[0000-0002-5286-2564]{Tie Liu}
\affiliation{Shanghai Astronomical Observatory, Chinese Academy of Sciences, 80 Nandan Road, Shanghai 200030, People’s Republic of China}

\author[0000-0003-3343-9645]{Hong-Li Liu}
\affiliation{School of Physics and Astronomy, Yunnan University, Kunming 650091, People’s Republic of China}

\author[0000-0003-4603-7119]{Sheng-Yuan Liu}
\affiliation{Academia Sinica Institute of Astronomy and Astrophysics, 11F AS/NTU Astronomy-Mathematics Building, No.1, Section 4, Roosevelt Road, Taipei 10617, Taiwan}

\author[0000-0002-5789-7504]{Meizhu Liu}
\affiliation{School of Physics and Astronomy, Yunnan University, Kunming 650091, People’s Republic of China}

\author[0000-0001-8277-1367]{Hongqiong Shi}
\affiliation{School of Physics and Astronomy, Yunnan University, Kunming 650091, People’s Republic of China}

\author[0000-0001-5710-6509]{Chuanshou Li}
\affiliation{School of Physics and Astronomy, Yunnan University, Kunming 650091, People’s Republic of China}

\author[0000-0001-9160-2944]{Mengyao Tang}
\affiliation{Institute of Astrophysics, School of Physics and Electronic Science, Chuxiong Normal University, Chuxiong 675000, People’s Republic of China}

\author[0000-0002-1466-3484]{Tianwei Zhang}
\affiliation{Research Center for Intelligent Computing Platforms, Zhejiang Laboratory, Hangzhou 311100, P.R.China}
\affiliation{I. Physikalisches Institut, Universit{\"a}t zu K{\"o}ln, Z{\"u}lpicher Stra{\ss}e 77, 50937 K{\"o}ln, Germany}

\author[0000-0002-8149-8546]{Ken'ichi Tatematsu}
\affiliation{National Astronomical Observatory of Japan, National Institutes of Natural Sciences, 2-21-1 Osawa, Mitaka, Tokyo 181-8588, Japan}

\author[0000-0003-2090-5416]{Xiaohu Li}
\affiliation{Xinjiang Astronomical Observatory, Chinese Academy of Sciences, Urumqi, China}

\author[0000-0001-5950-1932]{Fengwei Xu}
\affiliation{Kavli Institute for Astronomy and Astrophysics, Peking University, Beijing 100871, People’s Republic of China}
\affiliation{Department of Astronomy, School of Physics, Peking University, Beijing 100871, People's Republic of China}

\author[0000-0002-5076-7520]{Yuefang Wu}
\affiliation{Department of Astronomy, School of Physics, Peking University, Beijing 100871, People's Republic of China}

\author[0009-0004-6159-5375]{Dongting Yang}
\affiliation{School of Physics and Astronomy, Yunnan University, Kunming 650091, People’s Republic of China}

\begin{abstract}

Investigating the temperature and density structures of gas in massive protoclusters is crucial for understanding the chemical properties therein. In this study, we present observations of the continuum and thioformaldehyde (H$_{2}$CS) lines at 345 GHz of 11 massive protoclusters using the Atacama Large Millimeter/submillimeter Array (ALMA) telescope. High spatial resolution and sensitivity observations have detected 145 continuum cores from the 11 sources. H$_{2}$CS line transitions are observed in 72 out of 145 cores, including line-rich cores, warm cores and cold cores. The H$_{2}$ column densities of the 72 cores are estimated from the continuum emission which are larger than the density threshold value for star formation, suggesting that H$_{2}$CS can be widely distributed in star-forming cores with different physical environments. Rotation temperature and column density of H$_{2}$CS are derived by use of the XCLASS software. The results show the H$_{2}$CS abundances increase as temperature rises and higher gas temperatures are usually associated with higher H$_{2}$CS column densities. The abundances of H$_{2}$CS are positively correlated with its column density, suggesting that the H$_{2}$CS abundances are enhanced from cold cores, warm cores to line-rich cores in star forming regions.

\end{abstract}


\section{Introduction} \label{sec:intro}

The origin and evolution of life in the universe is one of the most fascinating and challenging questions in science. To address this question, astronomers have been searching for and studying the molecules that contain basic elements essential for life, such as hydrogen, carbon, nitrogen, oxygen, and sulfur. These elements form a variety of complex organic molecules in interstellar space, some of which may be precursors or building blocks of life \citep{1987ASSL..134..561I}. 

As of September 2023, more than 300 interstellar or circumstellar species have been discovered in space and their information is collected in the Cologne Database for Molecular Spectroscopy (CDMS, \citealt{2001A&A...370L..49M, 2005JMoSt.742..215M, 2016JMoSp.327...95E}). Surprisingly, even though sulfur is the tenth most abundant element in the universe \citep{2017iace.book.....Y}, there is a significant number (35) of sulfur-bearing molecules. The observations highlight the active role that sulfur plays in various interstellar environments, as sulfur plays a crucial role in the synthesis and evolution of macromolecules such as amino acids, proteins, and nucleic acids \citep{2018SciA....4.3302R}. One of the simplest and most abundant sulfur-containing molecules in space is thioformaldehyde (H$_{2}$CS), which has been detected in out Galaxy and in various environments in our Galaxy, ranging from  massive star-forming regions to low-mass protostars to comets \citep{1973AuJPh..26...85S, 2000DPS....32.4402W, 2011ApJ...737L..25M, 2015A&A...583A...1L, 2016ApJ...824...88O, 2016ApJ...827...72S, 2018IAUS..332...73O, 2019A&A...632A..83S, 2022ApJ...931..102S}. These observations suggest that H$_{2}$CS may be involved in prebiotic chemistry and may provide clues about the origin of life in the universe \citep{1992ApJS...82..167H, 2019ApJ...878...64H}.

Like its oxygen-substituted analog H$_{2}$CO, the structure of H$_{2}$CS is a nearly elongated and slightly asymmetrical rotor that exhibits a wealth of transitions in the millimeter and submillimeter bands \citep{2005JChPh.123o4310B,2010A&A...517A..96T,2013ApJ...771...95O,2014ApJ...789....8N,2019A&A...625A.147A,2022A&A...661A.111S}, and these spectral lines correspond to different energy levels. Due to the fact that its transitions between multiple levels at K$_{p}$ ladders can be captured within a single sideband spectrum \citep{1994ApJ...428..680B}, the physical parameters of celestial environments can be easily and accurately determined by fitting multiple lines of multiple species within a single spectral band. The transitions of H$_{2}$CS molecule are affected by the excitation of collisions with other molecules \citep{1993ApJS...89..123M,2009ASPC..417..219W,2012cosp...39..305C,2016ApJ...824...88O,2017NewA...52...48S}. In high-density environments, the impact of collision excitation on energy-level transitions is more significant, so the spectral lines of H$_{2}$CS can be used to study the physical conditions in dense molecular clouds. Because the emission of H$_{2}$CS at millimeter and submillimeter wavebands is optically thin \citep{2019ApJ...885...82L,2020ApJ...901...62L}, H$_{2}$CS is a better tracer than H$_{2}$CO for identifying the kinetic temperature and density of dense gases \citep{2017A&A...598A..30T,2020A&A...633A...7V}.

In this paper, we present the first systematical analysis of H$_{2}$CS molecule in 11 massive protoclusters using Band-7 data from cycle 5 of the Atacama Large Millimeter/submillimeter Array (ALMA) telescope. We aim to investigate the physical and chemical properties of H$_{2}$CS in these protoclusters and to explore its potential as a tracer of dense gas. We used multiple transitions of H$_{2}$CS at different energy levels to derive the kinetic temperature, density, column density, and abundance of H$_{2}$CS in our sample. The paper is organized as follows: In Section \ref{obs}, we describe our observations and data reduction. In Section \ref{res}, we present our results on the detection and distribution of H$_{2}$CS. In Section \ref{dis}, we analyze the physical and chemical parameters of H$_{2}$CS. In Section \ref{con}, we summarize our main conclusions.

\section{Observations}
\label{obs}

\subsection{Sample Selection}

The targeted 11 IRAS sources are a subsample of the ATOMS survey toward 146 massive protoclusters  by \cite{2016ApJ...829...59L}. The 146 massive protoclusters are very diverse that are suitable for statistics studies for the physical and chemical processes of star formation with different physical conditions. Among them, 30 sources are identified with "blue profiles" which indicate infall emotions \citep{2014MNRAS.444..874C}. Further more, 18 out of the 30 sources are found to be featured by HCN(3-2) and CO(4-3) lines and have virial parameters less than 2 \citep{2021RAA....21...14Y}, suggesting that the 18 sources are experiencing global collapse and undergoing star formation. Finally, 11 massive and luminous sources with collapsing signature were observed with the ALMA.

Table~\ref{tab:sample} shows key parameters of our sample, such as J2000.0 positions, systemic velocities, distances from the sun and Galactic center, source radius, averaged dust temperatures, bolometric luminosity, and clump mass. Our sources are high-mass star forming regions with bright CS $\rm J = 2-1$ emission (T$_b$ $>$ 2 K) indicating dense gas \citep{1996A&AS..115...81B}, and bolometric luminosities above 10$^4$ L$_{\sun}$ implying at least a B0.5 type star \citep{2004A&A...426...97F}. The gas mass of all sources exceeds 10$^3$ M$_{\sun}$. The clump-averaged temperatures range from 23 to 32 K with a median of 28 K.

\subsection{ALMA Observations and data reduction}

We observed 11 massive protocluster clumps with ALMA cycle 5 Band-7 from May 18 to May 20, 2018 (UTC) using 43 12-meter antennas in the C43-1 configuration (Project ID: 2017.1.00545.S; PI: Tie Liu). The observations covered four spectral windows (SPWs 31, 29, 27 and 25) with increasing spectral frequency ranges: (i) 342.36-344.24 GHz, (ii) 344.25-346.09 GHz, (iii) 354.27-354.74  GHz, and (iv) 356.60-357.07 GHz. Each window has 1920 channels with a bandwidth of 1875.00 MHz for SPW 31 and SPW 29, and 468.75 MHz for SPW 27 and SPW 25. The integrate on-source time was $\sim$ 3.7 minutes per source. J1650-5044 and J1924-2914 were used as atmosphere, flux and phase calibrators, while J1924-2914 as a bandpass calibrator. We calibrated the visibility data set using CASA software version 5.1.15 with the standard pipeline provided by the ALMA Observatory and imaged it with the TCLEAN task in CASA 5.3.

ALMA Band-7 is ideal for studying interstellar molecules, especially organic and inorganic ones, at high resolution and sensitivity \citep{2019A&A...629A..29J,2021A&A...645A..53M,2023arXiv230616959L}. We analyzed the spectral lines in two spectral windows (SPW 31 and SPW 29) because of their wide frequency band, which contain a rich molecular line transitions. With a spatial resolution of $\sim$\,0.8--1.2$\arcsec$, and a high sensitivity of $\sim$\,1.2 mJy beam$^{-1}$ for continuum and $\sim$\,4.7 mJy per channel for lines, we could easily observe molecular emission lines that could not be found in previous lower resolution observations \citep{2023ApJ...946L..41S}.

\begin{deluxetable*}{ccccccccccc}
\tabletypesize{\scriptsize}
\tablewidth{0pt} 
%\tablenum{1}
\tablecaption{Physics parameters of the targets in the observations.
\label{tab:sample}}
\tablehead{
\colhead{ID} & \colhead{IRAS}& \colhead{RA} & \colhead{DEC} & \colhead{V$_{\rm lsr}$} & \colhead{Distance} & \colhead{$\rm R_{\rm GC}$} & \colhead{Radius} & \colhead{$\rm T_{\rm dust}$} & \colhead{log($\rm L_{\rm bol}$)} & \colhead{log($\rm M_{\rm clump}$)} \\
\colhead{} & \colhead{}& \colhead{} & \colhead{} & \colhead{(km~s$^{-1}$)} & \colhead{(kpc)} & \colhead{(kpc)} & \colhead{(pc)} & \colhead{(K)} & \colhead{(L$_{\sun}$)} & \colhead{(M$_{\sun}$)}
}
\colnumbers
\startdata
1 & I14382-6017	&  14:42:02.76 &	$-$60:30:35.1 & $-$60.7  & 7.7   & 6.0    & 1.68   & 28.0   &  5.2  & 3.6    \\
2 & I14498-5856	&  14:53:42.81 &	$-$59:08:56.5 & $-$49.3  & 3.2   & 6.4    & 0.74   & 26.7   &  4.4  & 3.0    \\
3 & I15520-5234	&  15:55:48.84 &	$-$52:43:06.2 & $-$41.3  & 2.7   & 6.2    & 0.67   & 32.2   &  5.1  & 3.2    \\
4 & I15596-5301	&  16:03:32.29 &	$-$53:09:28.1 & $-$72.1  & 10.1   & 5.2    & 1.81   & 28.5   &  5.5  & 3.9    \\
5 & I16060-5146	&  16:09:52.85 &	$-$51:54:54.7 & $-$91.6  & 5.3   & 4.5    & 1.24   & 32.2   &  5.8  & 3.9     \\
6 & I16071-5142	&  16:11:00.01 &	$-$51:50:21.6 & $-$87.0  & 5.3   & 4.5    & 1.21   & 23.9   &  4.8  & 3.7     \\
7 & I16076-5134	&  16:11:27.12 &	$-$51:41:56.9 & $-$87.7  & 5.3   & 4.5    & 1.57   & 30.1   &  5.3  & 3.6     \\
8 & I16272-4837	&  16:30:59.08 &	$-$48:43:53.3 & $-$46.6  & 2.9   & 5.8    & 0.84   & 23.1   &  4.3  & 3.2     \\
9 & I16351-4722	&  16:38:50.98 &	$-$47:27:57.8 & $-$41.4  & 3.0   & 5.7    & 0.69   & 30.4   &  4.9  & 3.2     \\
10 & I17204-3636 &  17:23:50.32 &	$-$36:38:58.1 & $-$18.2  & 3.3   & 5.1    & 0.60   & 25.8   &  4.2  & 2.9     \\
11 & I17220-3609 &  17:25:24.99 &   $-$36:12:45.1 & $-$93.7  & 8.0   & 1.3    & 2.41   & 25.4   &  5.7  & 4.3     \\
\enddata
\tablecomments{The 11 IRAS sources are list in the table with detailed physical parameters \citep{2018MNRAS.473.1059U, 2020MNRAS.496.2790L}. The system velocities of the sources in column (5) are measured from molecular line observations (e.g. CO, NH3, CS, etc.). The distances to all of the sources in column (6) and (7) are determined using a combination HI analysis, maser parallax and spectroscopic measurements. The radii of the sources in column (8) are calculated using their effective angular radii and the distances. The averaged dust temperatures in column (9) and bolometric luminosities in column (10) are derived from the SED fits. The clump masses in column (11) are estimated using the \cite{1983QJRAS..24..267H} method.}
\end{deluxetable*}

\section{RESULTS and analysis}
\label{res}

\subsection{Continuum}

\subsubsection{Core Identification}

We analyzed the flux density in continuum maps to establish appropriate contours that minimize noise effects on dust core identification, such as spurious voids in continuum emission that would be treated as closed contours for dense cores. In addition, the dust cores are required to have at least two closed contours above a 8 rms level. Following this, we identified 145 dust cores across 11 sources.

\begin{figure*}
\centering
\includegraphics[width=0.7\linewidth]{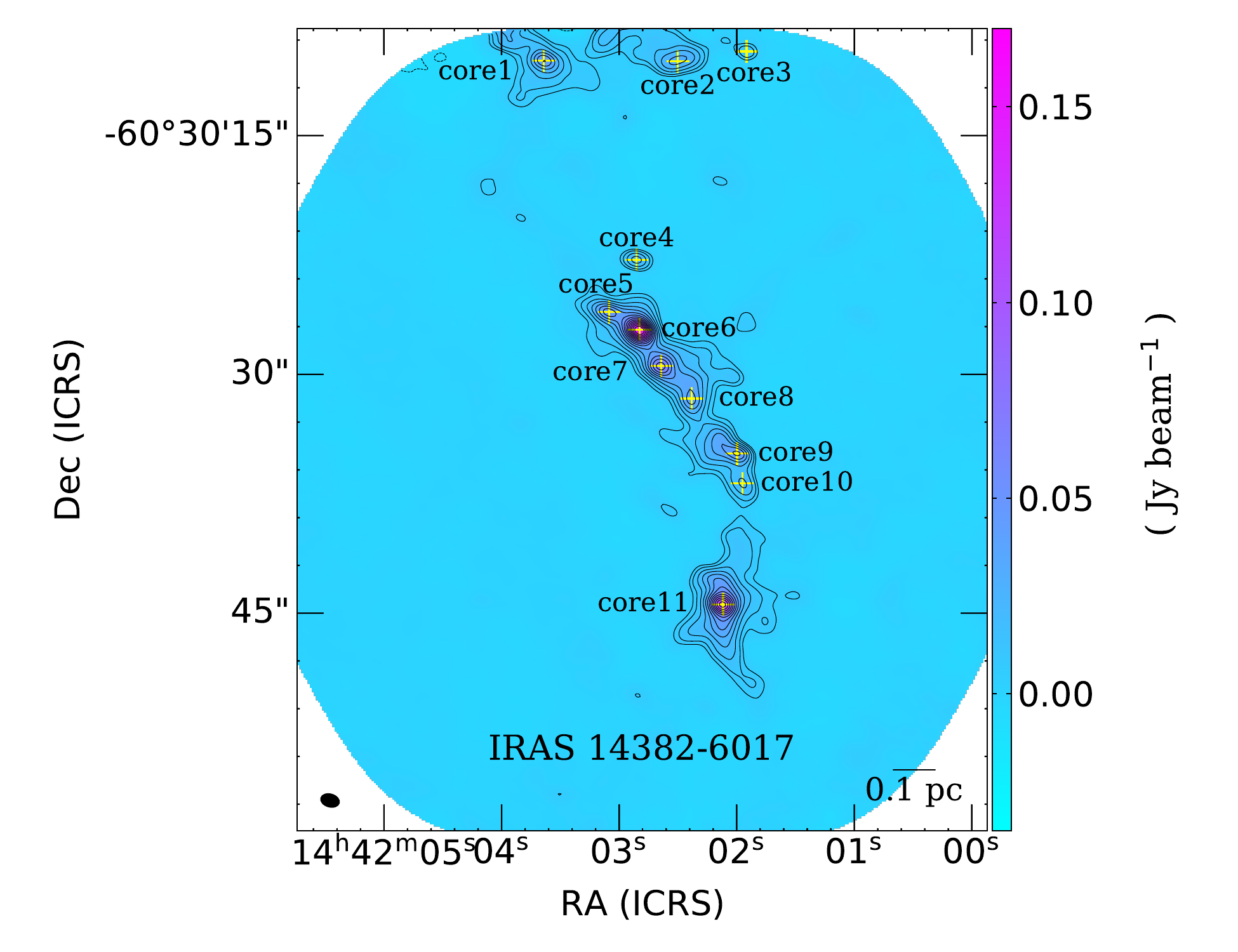}
\caption{345 GHz continuum maps of I14382 overlaid with the flux contours start with 8 rms and increase with the following power-law function $D=4\times N^{p}+8$, where $D$ is the dynamical range of the intensity map, $N$ is the number of contours used (15 in this case) and $p=Log(V_{max}/(4\times rms))/Log(N-1)$ (where $V_{max}$ is the peak intensity of the continuum) is the power index. The magnitude of the noise level of the continuum map is 0.8 mJy beam$^{-1}$. The cores is labeled roughly from top to bottom and from left to right. The corresponding beam size is in the bottom left corner and the scalebar is in the bottom right corner.}
\label{cores_id0}
\end{figure*}

We take the source IRAS 14382-6017 (hereafter I14382, as with other sources for the alias naming rule) as an example, which is shown in Figure~\ref{cores_id0}. The rms noise level is 0.8 mJy beam$^{-1}$ for continuum emission. To avoid noise effects, we set the contours to start at 8 rms with a 4 rms step. This results in clean contours around compact dust cores, allowing to locate them. As a result, 11 cores were identified in this source, as marked in the figure. Their peak intensities range from 13 to 142 mJy beam$^{-1}$. The core identification results of other 10 sources are shown in Figure~\ref{cores_id1} of the appendix.

\subsubsection{Core Parameters}\label{CP}

The deconvolved parameters of these dust cores were obtained using the 2D Gaussian fitting tool in CASA, including the full width at half maximum (FWHM) of the major and minor axes, denoted as $\rm \theta_{maj}$ and $\rm \theta_{min}$ respectively, along with the position angle (PA), integrated flux density, and peak intensity. These values are displayed in columns (6), (7), (9), and (10) of Table~\ref{tab:Gaussfit}. In regions with crowded cores, the CASA-IMFIT program separated them and made the accurate measurements for the parameters mentioned above. 

The mass of each core can be calculated as follows \citep{1983QJRAS..24..267H,2021MNRAS.505.2801L}:

\begin{equation}
 \label{eq:core_mass}
 M_{\text{core}} = \frac{D^2 S_\nu \eta}{\kappa_\nu B_\nu (T_d)},
\end{equation}
where D is the distance to the source, $S\rm _\nu$ is the integrated flux of the continuum, $\eta=100$ is a generic gas-to-dust ratio for interstellar matter with solar metallicity \citep{1991ApJ...380..429L, 1992ApJS...82..167H}, $\kappa_\nu=$ 1.89 cm$^2$ g$^{-1}$ is the dust absorption coefficient of molecular cloud cores at 870 $\mu$m \citep{1994A&A...291..943O}, and $\rm B_\nu (T_d)$ is the Planck function at the dust temperature $\rm T_d$. The method for selecting the dust temperature $\rm T_d$ is described in Section \ref{TS} in this article. The core with a larger distance tends to have a larger mass at the same integrated flux.

We list the core masses in Table~\ref{tab:Gaussfit} column (8). They range from 0.3 to 263.1 M${\sun}$. We find 75 massive cores ($>$ 8 M${\sun}$), 57 intermediate-mass cores (2$-$8 M${\sun}$), and 13 low-mass cores ($<$ 2 M${\sun}$). I17220 and I16060 have only massive cores, and I15596 has 21 massive cores. These are the three most massive sources of the 11 targets. No massive core is detected in I17204 and it is the lowest-mass source in the sample investigated here.

The mean optical depth of the continuum can be derived by the following equation \citep{2010ApJ...723.1665F,2021A&A...648A..66G}:

\begin{equation}
\textit{$\tau_{\nu}$}=-ln[1-\frac{{S_\nu}}{\Omega B_{\nu}(T_d)}]
\end{equation}
where $\Omega$ is the solid angle subtended by the source. The derived $\tau_{\nu}$ toward the 145 cores are on the order of 10$^{-3}$ to 10$^{-2}$, guaranteeing that optically thin assumption of dust continuum at 870 $\mu$m is reasonable. Then the source-averaged column density of H$_2$ ($\rm N_{H_2}$) can be derived as below \citep{2010ApJ...723.1665F, 2019A&A...628A..27B}:

\begin{equation}
 \label{eq:h2_cd}
 N_{\text{$H_2$}} = \frac{M_{\text{core}}}{\mu m_H \Omega D^2} 
                  = \frac{S_\nu \eta}{\mu m_H \Omega  \kappa_\nu B_\nu (T_d)},
\end{equation}
where $\mu\approx2.8$ is the mean particle weight per $\rm H_2$ molecule \citep{2008A&A...487..993K} and $\rm m_H$ is the hydrogen atom mass. Note that H$_2$CS abundances are not affected by distance which are derived from the ratios of source-averaged column densities of H$_2$CS and H$_2$.

We calculate $\rm N_{H_2}$ for the cores with H$_2$CS detected. They range from $3.1\times10^{22}$ to $4.1\times10^{24}$ $\rm cm^{-2}$, with a two orders of magnitude difference. Overall, massive cores have higher $\rm N_{H_2}$. All of the cores have column densities much above a threshold of $\sim7\times10^{21}$ $\rm cm^{-2}$ for $\rm N_{H_2}$ suggestive of initial density conditions of star formation \citep{2014prpl.conf...27A,2018ApJ...856..141T}, confirming that the protocluster sources investigated here are actively forming stars \citep{2020ApJ...890...44B}.

\subsection{H\texorpdfstring{$_2$}CCS Line Emission}\label{LE}

\begin{figure}[ht!]
\centering
\includegraphics[width=\linewidth]{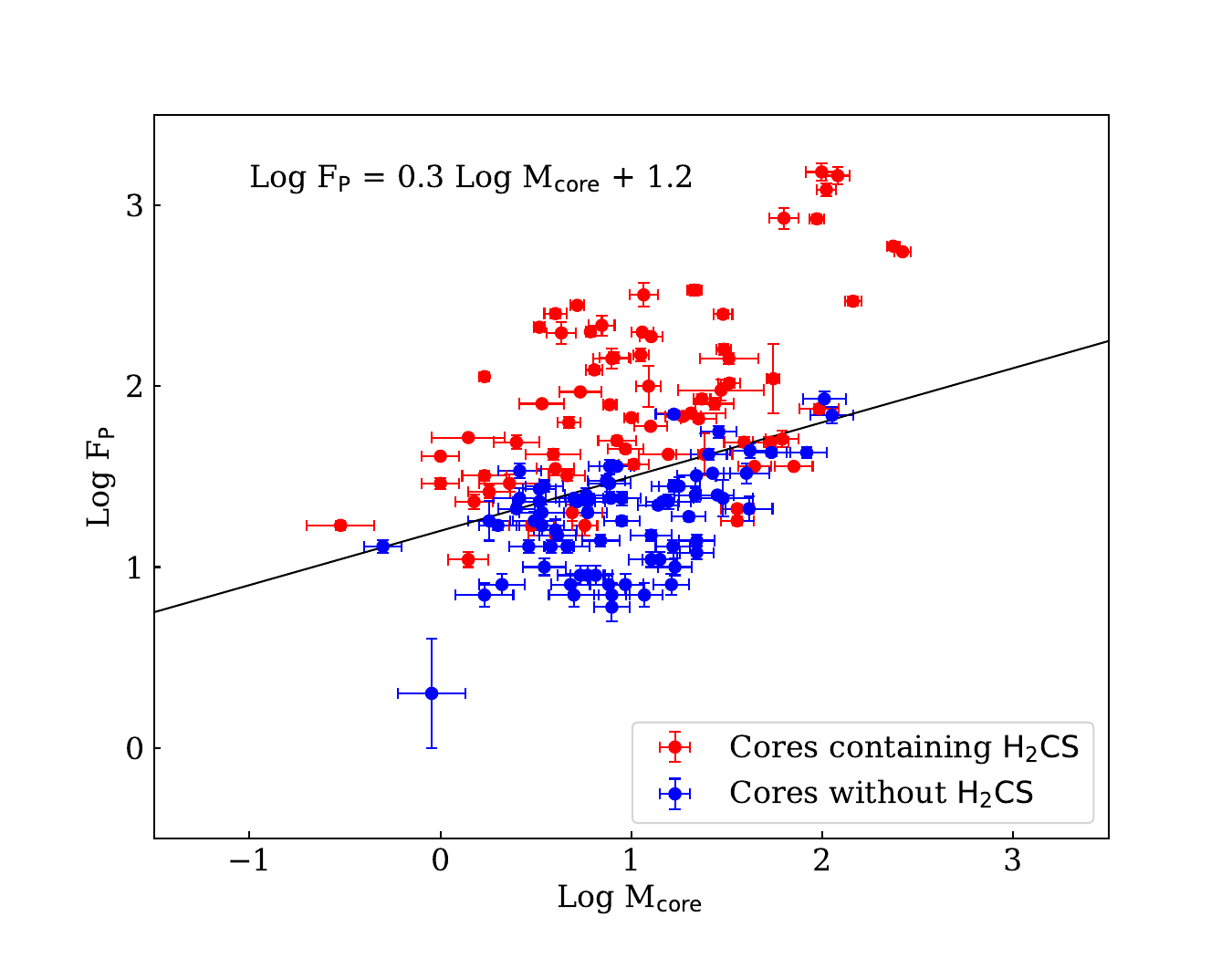}
\caption{The relationship between peak flux and mass of 145 cores, where red dots indicate cores where H$_{2}$CS is detected and blue dots indicate cores where H$_{2}$CS is not detected. The black slash marks a sharp distinction between the two types of cores.}
\label{Fp-M}
\end{figure}

Molecular lines are powerful tools for revealing the physical state of gas density structures and for exploring various astrochemical processes therein. As mentioned earlier, a total of 145 dense cores have been identified in 11 massive protocluster sources. We extracted the molecular lines from the continuum peak position of each core. Nine transitions of H$_{2}$CS are tuned in SPWs 29 and 31 \citep{2008ApJS..176..543M, 2019A&A...621A.143M}, and their corresponding parameters are summarized in Table~\ref{tab:H2CS}. The E$_{u}$ of H$_{2}$CS covers a wide range of energy levels from 91 to 419 K, allowing accurately determine the rotation temperature and column density. Therefore, two  isolated H$_{2}$CS lines with E$_{u}$ of 90.59 and 143.37 K and two mutually blended H$_{2}$CS lines with E$_{u}$ of 209.09 K were selected for a reliable parameters fitting, as they are separated from other molecular lines and spectrally resolved with higher signal to noise ratios. 

By inputting presupposed parameters (the deconvolved size of the continuum core, rotation temperature, source-averaged column density, FWHM, and $\rm V_{off}$), we utilized the XCLASS \citep{2017A&A...598A...7M} software to obtain the best-fit parameters for rotation temperature, source-averaged H$_{2}$CS column density, FWHM, and $\rm V_{off}$ for cores with $\geq$ 3 H$_{2}$CS lines. The $\rm V_{off}$ range was approximately determined relative to the systemic velocity $\rm V_{lsr}$ of each source, while the range of the other three parameters was constrained through manual matching. For cores with two velocity components, we set two H$_{2}$CS components with different parameters for simultaneous fitting, the final column density is the sum of the two components, and the temperature of the velocity component with higher temperature is used to calculate the column density of H$_{2}$. And for the cores with only one $\rm J_{K_a,K_c}=10_{0,10}-9_{0,9}$ line, the temperature was set to the same as that of the natal clump (with an error of 20\%), and subsequently the column density parameter was fitted separately.

We also calculated the optimal outcomes using two algorithm namely Genetic algorithm (GA) and Levenberg-Marquardt algorithm (LM) and determined the errors of temperatures and column densities using the Markov chain Monte Carlo algorithm (MCMC). All fitted H$_{2}$CS parameters are summarized in columns (5) to (8) of Table~\ref{tab:H2CSfit}.

Here we define line-rich cores as those with abundant molecular spectral lines (at least 50 lines) and potential CH$_{3}$OCHO detection \citep{2000prpl.conf..299K,2004ASPC..323..195C,2020ARA&A..58..727J}. Warm cores have H$_{2}$CS detection but no CH$_{3}$OCHO detection. Cold cores have only one or no H$_{2}$CS line. Among the 145 cores, 28 cores are considered line-rich with at least 50 lines detected in SPWs 29 and 31 \citep{2023ApJ...958..174L}. Additionally, 25 less-line-rich cores (warm cores) and 92 cores with several molecular lines (cold cores) are also found. We used the XCLASS \citep{2017A&A...598A...7M} software for H$_{2}$CS line identification. Among the 145 cores, 72 have H$_{\rm 2}$CS line emission. Multiple H$_{2}$CS lines can be detected in all line-rich and warm cores, and 19 out of 92 cold cores only have single H$_{2}$CS line transition. The reason whether a core has H$_{2}$CS transition detected may be attributed to the relationship between the continuum peak flux (F$\rm{_P}$) and mass of the core (M$\rm{_{core}}$). Figure~\ref{Fp-M} illustrates 145 cores that the peak flux of 84.7\% (61 out of 72) cores containing H$_{2}$CS meets Log F$\rm{_{P}}$ $>$ 0.3 Log M$\rm{_{core}}$ + 1.2 and the peak flux of 84.9\% (62 out of 73) cores without H$_{2}$CS has Log F$\rm{_{P}}$ $<$ 0.3 Log M$\rm{_{core}}$ + 1.2.

\begin{figure*}
\centering
\includegraphics[width=\linewidth]{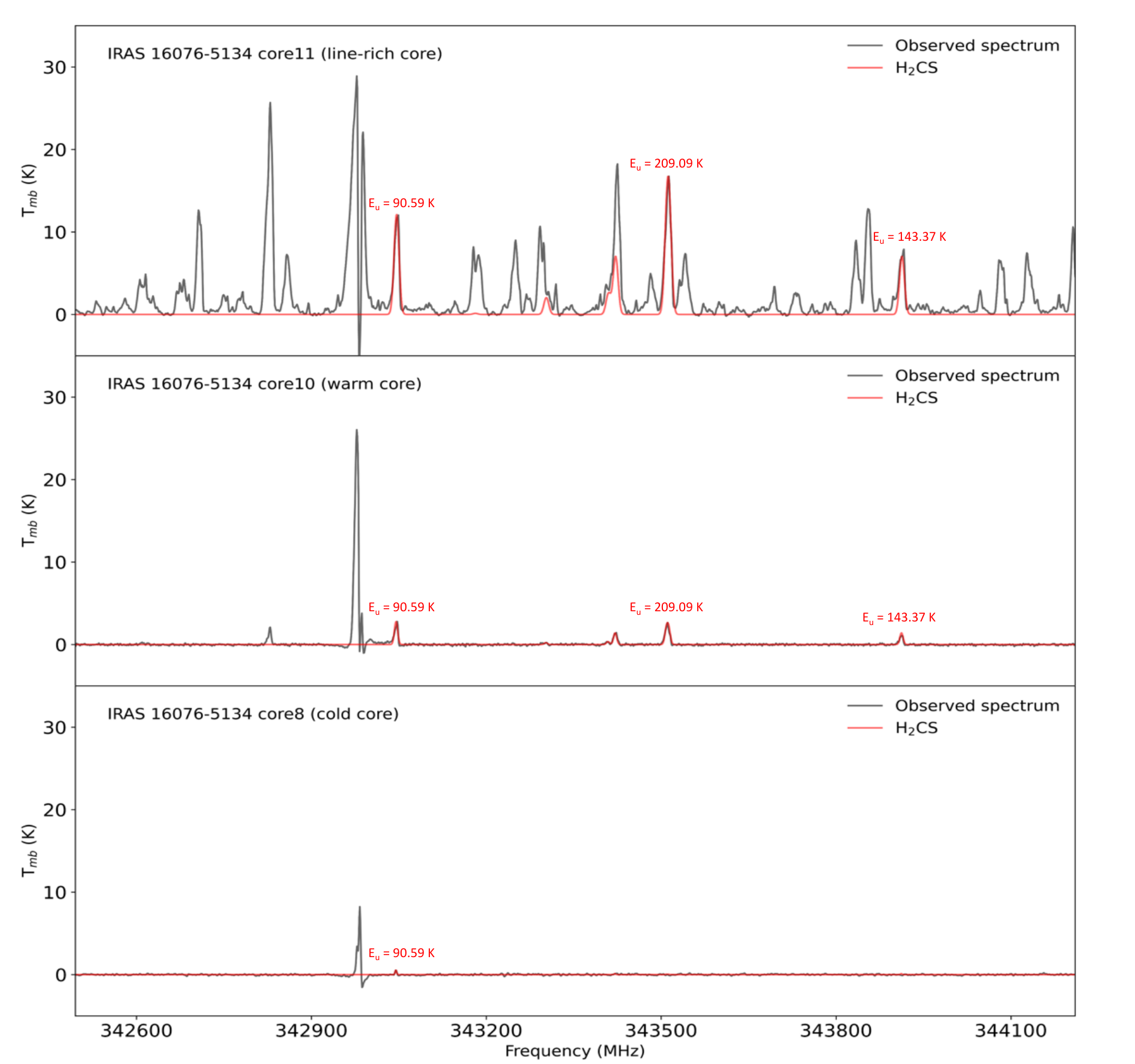}
\caption{Example of I16076 spectra encountered from the ALMA Band-7 survey. The black lines are observed spectra at sky frequencies and the red lines are synthetic spectra of the best fitting parameters for H$_{2}$CS lines. The three typical spectra of ``line-rich core", ``warm core" and ``cold core" are characterized by evidently rich, modest, and scarce of molecular lines. Compared with these three cores, owing to more molecular emission lines and wider line widths, core 11 has enhanced line confusion. Since the rms noise level in the spectrum is about 4.7 mJy per channel, most of the features seen are unambiguous line emission. }
\label{16076spec}
\end{figure*}

Figure~\ref{16076spec} shows typical H$_{2}$CS lines in I16076 for "line-rich core", "warm core", and "cold core". These three types of cores show different number and emission intensity of molecular lines. The fitting results of the other cores are shown in Figure~\ref{f:spec1}.

\startlongtable
\begin{deluxetable*}{cccccccccccc}
\tabletypesize{\scriptsize}
\tablewidth{0pt}
\tablecaption{Parameters of fitted H$_{\rm 2}$CS molecules among detected cores\label{tab:H2CSfit}}
\tablehead{
\colhead{ID} & \colhead{IRAS}& \colhead{Core} & \colhead{$\rm \theta_{\rm deconv}$} & \colhead{$\rm T_{\rm H_2CS}$} & \colhead{$\rm N_{H_2CS}$} & \colhead{FWHM} & \colhead{$\rm V_{\rm off}$} & \colhead{Type\tablenotemark{\footnotesize a}} & \colhead{$\rm T_{\rm CH_3OCHO}$} & \colhead{$\rm N_{H_2}$} & \colhead{$\rm f_{H_2CS}$} \\
\colhead{} & \colhead{}& \colhead{} & \colhead{(\arcsec)} & \colhead{(K)} & \colhead{($\rm cm^{\rm -2}$)} & \colhead{(km $\rm s^{\rm -1}$)} & \colhead{(km $\rm s^{\rm -1}$)} & \colhead{} & \colhead{(K)} & \colhead{($\times10^{23}$\,$\rm cm^{-2}$)} & \colhead{}
} 
\colnumbers
\startdata
1 & I14382-6017 & 5 & 1.13 & 47±14 & (6.8±3.5)$\times10^{13}$ & 2.0 & $-$59.8 & H & & 0.1±0.0 & (6.2±3.2)$\times10^{-10}$ \\
2 & I14382-6017 & 6 & 1.22 & 56±16 & (1.6±0.3)$\times10^{14}$ & 2.6 & $-$59.2 & H & & 2.2±0.7 & (7.3±2.5)$\times10^{-10}$ \\
3 & I14382-6017 & 7 & 1.34 & 50±17 & (7.1±0.2)$\times10^{13}$ & 1.4 & $-$59.2 & H & & 1.2±0.4 & (6.1±2.1)$\times10^{-10}$ \\
4 & I14382-6017 & 8 & 1.48 & 50±15 & (5.5±1.1)$\times10^{13}$ & 3.0 & $-$59.7 & H & & 0.7±0.2 & (7.6±2.8)$\times10^{-10}$ \\
5 & I14382-6017 & 11 & 1.37 & 51±5 & (4.1±2.6)$\times10^{13}$ & 3.0 & $-$59.2 & H & & 1.8±0.2 & (2.3±1.5)$\times10^{-10}$ \\
6 & I14498-5856	& 2 & 2.47 & 109±4 & (7.0±0.1)$\times10^{15}$ & 6.1 & $-$51.3 & C & 102±6 & 2.1±0.2 & (3.4±0.3)$\times10^{-8}$ \\
7 & I14498-5856	& 4 & 1.55 & 26±5 & (1.1±0.2)$\times10^{14}$ & 3.3 & $-$48.8 & G & & 2.5±0.5 & (4.5±1.2)$\times10^{-10}$ \\
8 & I14498-5856	& 5 & 1.63 & 26±5 & (6.0±1.5)$\times10^{13}$ & 2.0 & $-$47.5 & G & & 1.9±0.4 & (3.2±1.0)$\times10^{-10}$ \\ 
9 & I15520-5234 & 1 & 2.22 & 60±16 & (1.0±0.2)$\times10^{14}$ & 2.4 & $-$40.3 & H & & 0.7±0.2 & (4.4±1.1)$\times10^{-9}$ \\
 & & & & 55±11 & (2.1±0.2)$\times10^{14}$ & 2.5 & $-$44.7 \\
10 & I15520-5234 & 2 & 0.75 & 60±14 & (1.3±0.2)$\times10^{14}$ & 2.5 & $-$46.2 & H & & 1.5±0.4 & (1.7±0.6)$\times10^{-9}$ \\
 & & & & 43±13 & (1.2±0.4)$\times10^{14}$ & 2.3 & $-$43.1 \\
11 & I15520-5234 & 3 & 1.61 & 50±8 & (1.1±0.2)$\times10^{14}$ & 3.0 & $-$43.7 & H & & 0.5±0.1 & (2.3±1.1)$\times10^{-9}$ \\
12 & I15520-5234 & 4 & 3.04 & 78±7 & (2.7±0.1)$\times10^{15}$ & 3.5 & $-$43.5 & C & 77±11 & 2.7±0.3 & (1.5±0.1)$\times10^{-8}$ \\
 & & & & 64±8 & (1.3±0.1)$\times10^{15}$ & 3.2 & $-$39.7 \\
13 & I15520-5234 & 5 & 1.21 & 42±10 & (3.1±1.5)$\times10^{14}$ & 2.3 & $-$42.0 & H & & 1.0±0.2 & (3.2±1.8)$\times10^{-9}$ \\
14 & I15520-5234 & 6 & 2.41 & 70±3 & (3.6±0.1)$\times10^{15}$ & 3.4 & $-$40.5 & C & 80±10 & 1.6±0.1 & (2.3±0.2)$\times10^{-8}$ \\
15 & I15520-5234 & 7 & 1.45 & 79±3 & (4.1±0.1)$\times10^{15}$ & 3.8 & $-$39.6 & C & 78±10 & 2.8±0.4 & (1.5±0.2)$\times10^{-8}$ \\
16 & I15520-5234 & 8 & 1.39 & 100±4 & (2.2±0.3)$\times10^{15}$ & 3.8 & $-$44.2 & C & 104±6 & 1.9±0.3 & (1.2±0.3)$\times10^{-8}$ \\
17 & I15520-5234 & 9 & 1.67 & 107±4 & (5.6±0.1)$\times10^{15}$ & 1.8 & $-$44.3 & C & 102±4 & 1.8±0.1 & (3.1±0.2)$\times10^{-8}$ \\
18 & I15520-5234 & 10 & 1.63 & 74±18 & (1.9±0.1)$\times10^{15}$ & 2.6 & $-$39.1 & C & 70±36 & 1.1±0.2 & (3.5±0.4)$\times10^{-8}$ \\
 & & & & 63±11 & (1.8±0.2)$\times10^{15}$ & 1.6 & $-$42.8 \\
19 & I15520-5234 & 11 & 2.43 & 75±12 & (1.6±0.2)$\times10^{15}$ & 2.7 & $-$42.9 & C & 105±37 & 1.1±0.2 & (1.4±0.3)$\times10^{-8}$ \\
20 & I15520-5234 & 12 & 1.91 & 62±14 & (3.0±1.3)$\times10^{14}$ & 2.4 & $-$42.5 & H & & 0.4±0.1 & (1.7±0.4)$\times10^{-8}$ \\
 & & & & 46±13 & (3.8±1.0)$\times10^{14}$ & 2.7 & $-$46.1 \\
21 & I15520-5234 & 13 & 2.20 & 60±16 & (2.3±0.5)$\times10^{14}$ & 3.0 & $-$41.0 & H & & 0.5±0.1 & (4.9±1.8)$\times10^{-9}$ \\
22 & I15520-5234 & 15 & 1.04 & 50±9 & (8.2±4.0)$\times10^{13}$ & 2.7 & $-$42.7 & H & & 0.8±0.2 & (1.1±0.6)$\times10^{-9}$ \\
23 & I15596-5301 & 11 & 1.19 & 28±5 & (8.4±1.8)$\times10^{13}$ & 2.8 & $-$72.9 & G & & 2.4±0.4 & (3.4±1.0)$\times10^{-10}$ \\
24 & I15596-5301 & 12 & 1.09 & 28±5 & (5.6±1.2)$\times10^{13}$ & 0.7 & $-$72.5 & G & & 2.5±0.5 & (2.2±0.6)$\times10^{-10}$ \\
25 & I15596-5301 & 13 & 1.02 & 80±8 & (1.7±0.2)$\times10^{15}$ & 4.5 & $-$73.0 & C & 100±23 & 0.7±0.1 & (2.4±0.5)$\times10^{-8}$ \\
26 & I15596-5301 & 15 & 1.17 & 28±5 & (1.7±0.4)$\times10^{14}$ & 1.2 & $-$71.6 & G & & 1.9±0.3 & (8.8±2.6)$\times10^{-10}$ \\
27 & I15596-5301 & 16 & 1.50 & 28±5 & (9.5±2.1)$\times10^{13}$ & 2.3 & $-$71.2 & G & & 0.9±0.2 & (1.0±0.3)$\times10^{-9}$ \\
28 & I15596-5301 & 17 & 1.73 & 85±8 & (1.5±0.1)$\times10^{15}$ & 3.0 & $-$77.7 & H & & 0.5±0.1 & (3.8±0.9)$\times10^{-8}$ \\
 & & & & 66±13 & (3.2±0.2)$\times10^{14}$ & 2.4 & $-$73.0 \\
29 & I15596-5301 & 18 & 1.41 & 99±3 & (4.7±0.5)$\times10^{15}$ & 5.2 & $-$71.3 & C & 110±39 & 1.7±0.1 & (2.8±0.4)$\times10^{-8}$ \\
30 & I15596-5301 & 20 & 0.70 & 50±18 & (5.0±0.1)$\times10^{14}$ & 4.5 & $-$74.9 & H & & 3.7±1.5 & (1.4±0.6)$\times10^{-9}$ \\
31 & I15596-5301 & 21 & 1.66 & 28±5 & (1.9±0.4)$\times10^{14}$ & 2.8 & $-$74.9 & G & & 0.8±0.1 & (2.5±0.7)$\times10^{-9}$ \\
32 & I16060-5146 & 2 & 1.33 & 32±6 & (5.8±1.0)$\times10^{13}$ & 1.5 & $-$93.4 & G & & 2.8±0.5 & (2.1±0.6)$\times10^{-10}$ \\
33 & I16060-5146 & 4 & 1.33 & 110±13 & (4.7±0.1)$\times10^{15}$ & 4.7 & $-$96.5 & C & 136±4 & 12.1±2.1 & (3.9±0.7)$\times10^{-9}$ \\
34 & I16060-5146 & 5 & 1.62 & 110±6 & (5.6±0.1)$\times10^{15}$ & 3.0 & $-$84.8 & C & 112±7 & 7.6±0.7 & (7.3±0.6)$\times10^{-9}$ \\
35 & I16060-5146 & 7 & 1.44 & 121±4 & (1.2±0.3)$\times10^{15}$ & 6.4 & $-$95.8 & H & & 12.5±1.6 & (9.6±2.7)$\times10^{-10}$ \\
36 & I16060-5146 & 8 & 1.39 & 93±4 & (1.6±0.1)$\times10^{15}$ & 6.4 & $-$86.3 & C & 110±2 & 11.8±1.3 & (1.4±0.2)$\times10^{-9}$ \\
37 & I16060-5146 & 9 & 0.92 & 86±11 & (2.0±0.1)$\times10^{15}$ & 4.4 & $-$92.9 & H & & 3.3±0.5 & (1.1±0.1)$\times10^{-8}$ \\
 & & & & 81±9 & (1.7±0.1)$\times10^{15}$ & 5.2 & $-$86.9 \\
38 & I16060-5146 & 10 & 0.94 & 32±6 & (9.8±1.7)$\times10^{13}$ & 1.5 & $-$87.8 & G & & 2.3±0.4 & (4.3±1.1)$\times10^{-10}$ \\
39 & I16060-5146 & 11 & 1.43 & 40±3 & (1.3±0.1)$\times10^{14}$ & 2.7 & $-$97.4 & H & & 2.5±0.2 & (5.3±0.7)$\times10^{-10}$ \\
40 & I16060-5146 & 12 & 1.10 & 32±6 & (2.7±0.5)$\times10^{14}$ & 2.5 & $-$94.3 & G & & 3.3±0.6 & (8.1±2.2)$\times10^{-10}$ \\
41 & I16060-5146 & 13 & 1.55 & 46±7 & (2.7±1.0)$\times10^{14}$ & 2.2 & $-$92.4 & H & & 0.9±0.2 & (2.9±1.2)$\times10^{-9}$ \\
42 & I16071-5142 & 5 & 0.59 & 23±4 & (4.5±1.3)$\times10^{14}$ & 3.5 & $-$84.4 & G & & 7.8±1.4 & (5.8±2.0)$\times10^{-10}$ \\
43 & I16071-5142 & 6 & 1.34 & 95±4 & (2.6±0.1)$\times10^{16}$ & 8.7 & $-$87.2 & C & 127±10 & 7.5±1.2 & (3.5±0.6)$\times10^{-8}$ \\
44 & I16076-5134 & 8 & 1.98 & 30±6 & (1.2±0.3)$\times10^{14}$ & 2.4 & $-$86.1 & G & & 2.1±0.5 & (5.6±1.8)$\times10^{-10}$ \\
45 & I16076-5134 & 9 & 0.94 & 81±18 & (4.5±1.5)$\times10^{14}$ & 3.2 & $-$89.5 & C & 87±11 & 1.9±0.4 & (2.4±1.0)$\times10^{-9}$ \\
46 & I16076-5134 & 10 & 2.30 & 81±8 & (8.9±1.0)$\times10^{14}$ & 5.6 & $-$86.4 & H & & 0.2±0.0 & (3.9±0.7)$\times10^{-8}$ \\
47 & I16076-5134 & 11 & 1.86 & 98±3 & (7.1±0.1)$\times10^{14}$ & 8.5 & $-$87.2 & C & 98±10 & 1.9±0.2 & (3.7±0.3)$\times10^{-9}$ \\
48 & I16076-5134 & 12 & 1.20 & 42±12 & (1.4±0.6)$\times10^{14}$ & 1.3 & $-$89.9 & H & & 0.7±0.2 & (1.9±1.0)$\times10^{-9}$ \\
49 & I16076-5134 & 14 & 0.92 & 30±6 & (2.8±0.6)$\times10^{14}$ & 3.5 & $-$87.2 & G & & 1.0±0.3 & (2.7±1.0)$\times10^{-9}$ \\
50 & I16076-5134 & 15 & 1.28 & 68±5 & (8.4±1.2)$\times10^{14}$ & 3.0 & $-$87.4 & H & & 1.3±0.1 & (6.4±1.1)$\times10^{-9}$ \\
51 & I16272-4837 & 4 & 0.89 & 89±3 & (1.5±0.1)$\times10^{16}$ & 3.2 & $-$46.6 & C & 106±3 & 3.0±0.2 & (5.1±0.5)$\times10^{-8}$ \\
52 & I16272-4837 & 5 & 1.16 & 23±4 & (8.5±2.5)$\times10^{13}$ & 2.0 & $-$45.5 & G & & 2.5±0.5 & (3.5±1.2)$\times10^{-10}$ \\
53 & I16272-4837 & 6 & 0.85 & 90±9 & (6.8±0.2)$\times10^{15}$ & 8.0 & $-$48.4 & C & 102±17 & 4.0±0.5 & (1.7±0.2)$\times10^{-8}$ \\
54 & I16272-4837 & 7 & 0.81 & 71±5 & (1.4±0.1)$\times10^{15}$ & 2.4 & $-$46.4 & C & 82±27 & 5.7±0.4 & (2.5±0.3)$\times10^{-9}$ \\
55 & I16272-4837 & 8 & 0.80 & 112±6 & (3.6±0.1)$\times10^{16}$ & 3.1 & $-$46.4 & C & 115±4 & 15.7±1.3 & (2.3±0.2)$\times10^{-8}$ \\
56 & I16351-4722 & 4 & 0.69 & 70±5 & (8.4±0.3)$\times10^{14}$ & 1.8 & $-$42.0 & C & 90±9 & 2.4±0.2 & (3.5±0.3)$\times10^{-9}$ \\
57 & I16351-4722 & 5 & 1.25 & 30±6 & (9.8±2.0)$\times10^{13}$ & 1.5 & $-$42.4 & G & & 1.8±0.4 & (5.5±1.6)$\times10^{-10}$ \\
58 & I16351-4722 & 6 & 1.74 & 87±12 & (3.0±0.2)$\times10^{15}$ & 4.8 & $-$43.3 & C & 70±4 & 1.8±0.3 & (2.0±0.3)$\times10^{-8}$ \\
 & & & & 76±18 & (5.3±0.2)$\times10^{14}$ & 2.2 & $-$38.1 \\
59 & I16351-4722 & 7 & 2.04 & 101±1 & (3.0±0.1)$\times10^{16}$ & 5.5 & $-$39.6 & C & 170±10 & 1.9±0.3 & (1.6±0.3)$\times10^{-7}$ \\
60 & I16351-4722 & 8 & 1.83 & 81±10 & (5.6±0.3)$\times10^{15}$ & 4.8 & $-$39.0 & C & 91±4 & 2.3±0.3 & (2.5±0.3)$\times10^{-8}$ \\
61 & I16351-4722 & 9 & 1.76 & 56±4 & (4.8±0.2)$\times10^{14}$ & 1.0 & $-$37.7 & H & & 1.6±0.1 & (3.0±0.3)$\times10^{-9}$ \\
62 & I16351-4722 & 10 & 1.33 & 58±7 & (4.2±0.8)$\times10^{14}$ & 4.0 & $-$38.1 & H & & 1.8±0.2 & (2.4±0.6)$\times10^{-9}$ \\
63 & I16351-4722 & 11 & 0.70 & 78±5 & (4.7±0.1)$\times10^{14}$ & 3.0 & $-$37.6 & H & & 0.4±0.0 & (1.1±0.1)$\times10^{-8}$ \\
64 & I16351-4722 & 12 & 1.10 & 51±12 & (3.4±1.1)$\times10^{14}$ & 3.9 & $-$40.1 & H & & 1.4±0.3 & (2.4±1.0)$\times10^{-9}$ \\
65 & I17204-3636 & 4 & 0.98 & 25±5 & (1.8±0.5)$\times10^{14}$ & 3.0 & $-$16.9 & G & & 0.8±0.2 & (2.2±0.8)$\times10^{-9}$ \\
66 & I17204-3636 & 5 & 1.28 & 25±5 & (7.0±2.0)$\times10^{13}$ & 2.0 & $-$16.9 & G & & 1.0±0.2 & (6.9±2.5)$\times10^{-10}$ \\
67 & I17204-3636 & 9 & 1.55 & 88±7 & (3.0±0.2)$\times10^{14}$ & 3.0 & $-$17.5 & H & & 1.5±0.1 & (2.0±0.2)$\times10^{-9}$ \\
68 & I17220-3609 & 3 & 1.48 & 25±5 & (2.4±0.7)$\times10^{14}$ & 2.5 & $-$94.1 & G & & 4.1±0.9 & (5.8±2.1)$\times10^{-10}$ \\
69 & I17220-3609 & 7 & 1.70 & 68±5 & (3.8±0.6)$\times10^{14}$ & 5.0 & $-$94.6 & H & & 4.8±0.5 & (8.0±1.5)$\times10^{-10}$ \\
70 & I17220-3609 & 9 & 2.24 & 100±9 & (2.5±0.1)$\times10^{16}$ & 7.2 & $-$96.1 & C & 107±13 & 4.9±0.5 & (5.1±0.5)$\times10^{-8}$ \\
71 & I17220-3609 & 10 & 1.84 & 117±6 & (2.4±1.2)$\times10^{16}$ & 7.4 & $-$96.7 & C & 94±3 & 6.6±0.5 & (3.7±1.8)$\times10^{-8}$ \\
72 & I17220-3609 & 14 & 1.91 & 25±5 & (1.5±0.4)$\times10^{14}$ & 3.0 & $-$95.1 & G & & 1.8±0.4 & (8.2±2.7)$\times10^{-10}$ \\
\enddata
\tablecomments{IRAS sources and extracted cores ID are listed in column (2) and (3). The calculated deconvolved source sizes $\theta$ are listed in column (4). The fitted rotational temperatures, column densities, FWHM, and velocity offset ($\rm V_{off}$) of H$_{2}$CS are listed in column (5)--(8), for 8 cores with two velocity components, the parameters are listed simultaneously. The fitted rotational temperatures of CH$_3$OCHO are list in column (10) (C. Li et al. 2023, submitted to ApJ). The peak column densities of H$_2$ are shown in column (11) and the abundances of H$_{2}$CS relative to H$_{2}$ are shown in column (12).}
\tablenotetext{a}{Column (9) presents the type of each core: G means neither H$_{2}$CS nor CH$_3$OCHO is detected; H means H$_{2}$CS is detected but CH$_3$OCHO is not detected; C means both H$_{2}$CS and CH$_3$OCHO are detected.}
\end{deluxetable*}

\subsection{H\texorpdfstring{$_2$}CCS Spatial Distribution}

\begin{figure*}[!ht]
\centering
\includegraphics[width=0.95\linewidth]{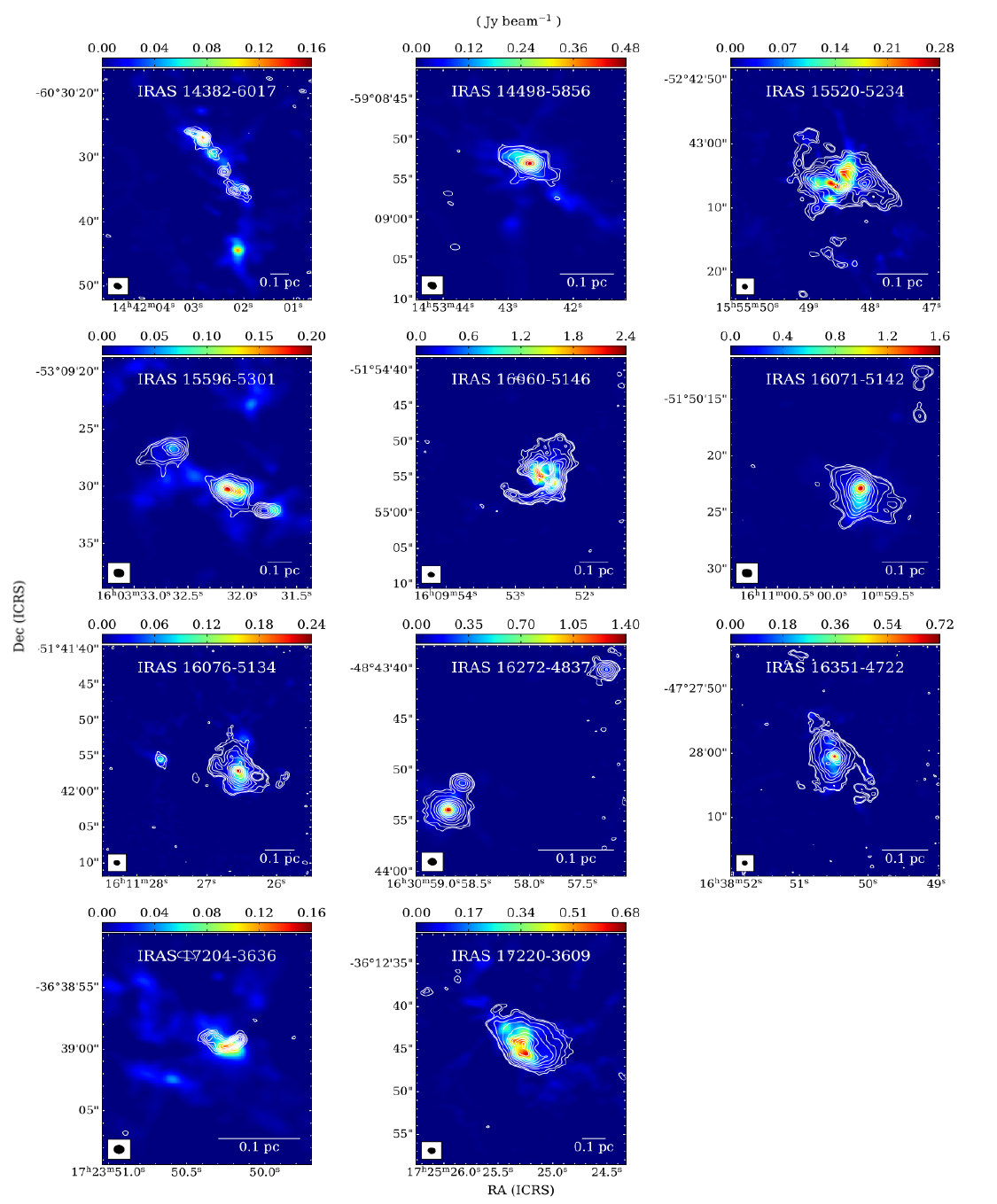}
\caption{345GHz continuum maps of the 11 sources overlaid with the Moment 0 contours of H$_{2}$CS $\rm J_{K_a,K_c}=10_{0,10}-9_{0,9}$ transition. The contours start with 4 rms and increase with the following power-law function $D=3\times N^{p}+2$ where $D$ is the dynamical range of the Moment 0 map, $N$ is the number of contours used (11 in this case) and $p=Log(V_{max}/(4\times rms))/Log(N-1)$ (where $V_{max}$ is the peak intensity of the Moment 0 map) is the power index. The magnitude of the noise level of the Moment 0 maps associated with each source is as follows: 17, 37, 55, 40, 60, 55, 35, 80, 40, 36 and 70 mJy/beam.km/s. The corresponding beam sizes are in the bottom left corner and the scalebars are in the bottom right corner.}
\label{f:moment0}
\end{figure*}

Figure~\ref{f:moment0} shows the 870 $\mu$m continuum maps of the 11 protocluster clumps, overlaid with contours of the H$_{2}$CS $\rm J_{K_a,K_c}=10_{0,10}-9_{0,9}$ Moment 0 (integrated intensity) maps. For the continuum maps, we zoom in on the central regions of the protoclusters to see the intensities of H$_{2}$CS emissions in detail. The moment 0 maps demonstrate that H$_{2}$CS is widely distributed throughout the clumps.

Among our targeted 11 sources, I14382 and I17204 are 2 sources with a few molecular emission lines and only contain warm cores and cold cores, and H$_{2}$CS is distributed within very small areas around warm cores. For the two sources I14498 and I15596, the emissions of H$_{2}$CS molecule are mainly around the compact regions of the line-rich cores. And for the remaining 7 sources, the spatial distribution of H$_{2}$CS emission is similar with the dust emission. The observations revealed that the emission of H$_{\rm 2}$CS is originated from extended regions surrounding both compact line-rich cores and warm cores and also can cover plenty of regions of the cold cores.

\subsection{Temperature Structure}\label{TS}

Temperature is an indicator of the interstellar chemical complexity and star formation process \citep{2017A&A...604A...6S,2020ApJ...895...86G,2020ApJ...901...31L,2022MNRAS.512.4419P}. It affects the mass estimation of dense gas. It also provides the main thermal pressure source within molecular gas \citep{2020SSRv..216...62R,2022ApJ...941..202R}, which counteracts gravity in the early stages of star formation. Thus, accurate temperature measurements enable an appropriate understanding of star formation mechanism. H$_{2}$CS can provide such an useful tracer of gas temperature.

Table~\ref{tab:numbers} lists the number of these three-type cores. The H$_{2}$CS temperature ranges from 23 to 121 K for all cores. The range is 68 to 121 K for line-rich cores and 40 to 88 K for warm cores. The temperatures for cold cores range from 23 to 32 K, which are equal to the clump temperatures. Compared to the temperatures of CH$_{3}$OCHO of line-rich cores, H$_{2}$CS shows no significant temperature differences within uncertainties, because the molecule is thermalized in line-rich core regions. Therefore, we choose the higher gas temperature of H$_{2}$CS and CH$_{3}$OCHO as $\rm T_d$. In addition, H$_{2}$CS is more widely distributed than CH$_{3}$OCHO in protoclusters and can applicably determine the temperature of cold cores, warm cores and line-rich cores, indicating that H$_{2}$CS is a more efficient temperature tracer as it can trace more extended regions than CH$_{3}$OCHO.

\begin{deluxetable}{c|ccc|c}
\tabletypesize{\scriptsize}
\tablewidth{0pt} 
\tablecaption{Numbers of three-type cores\label{tab:numbers}}
\tablehead{
\colhead{Sources} & \multicolumn{3}{c}{Cores} & \colhead{Total} \\
\cline{2-4}
\colhead{} & \colhead{Line-rich} & \colhead{Warm} & \colhead{Cold} & \colhead{}
}
\startdata 
IRAS 14382-6017 & 0 & 5 & 6 & 11 \\
IRAS 14498-5856 & 1 & 0 & 6 & 7 \\
IRAS 15520-5234 & 7 & 7 & 1 & 15 \\
IRAS 15596-5301 & 2 & 2 & 23 & 27 \\
IRAS 16060-5146 & 4 & 3 & 6 & 13 \\
IRAS 16071-5142 & 1 & 0 & 6 & 7 \\
IRAS 16076-5134 & 2 & 3 & 10 & 15 \\
IRAS 16272-4837 & 4 & 0 & 5 & 9 \\
IRAS 16351-4722 & 4 & 4 & 4 & 12 \\
IRAS 17204-3636 & 0 & 1 & 11 & 12 \\
IRAS 17220-3609 & 3 & 0 & 14 & 17 \\
\hline
Total & 28 & 25 & 92 & 145 \\
\enddata
\end{deluxetable}

\section{Discussion}
\label{dis}

\subsection{H\texorpdfstring{$_2$}CCS Column Densities and Abundances}

Based on Table~\ref{tab:H2CSfit} and Figure \ref{X_fig}, several significant characteristics can be found among line-rich, warm, and cold cores by comparing $\rm N_{H_2}$ and $\rm N_{H_2CS}$ parameters. For line-rich cores, $\rm N_{H_2CS}$ ranges from $3.8\times10^{14}$ to $3.6\times10^{16}$ $\rm cm^{-2}$, with a mean of $8.0\times10^{15}$ $\rm cm^{-2}$. For warm and cold cores, $\rm N_{H_2CS}$ ranges from $2.1\times10^{13}$ to $1.9\times10^{15}$ $\rm cm^{-2}$ and $5.6\times10^{13}$ to $4.5\times10^{14}$ $\rm cm^{-2}$, with means of $4.0\times10^{14}$ and $1.5\times10^{14}$ $\rm cm^{-2}$, respectively. The line-rich cores have higher $\rm N_{H_2CS}$ values than the other two types of cores. The warm cores have a wider range of $\rm N_{H_2CS}$ than the cold ones, but the values overlap, despite the mean $\rm N_{H_2CS}$ of warm cores being much higher (about twice) than that of cold cores. 

The abundance of H$_2$CS versus H$_2$ ($\rm f_{H_2CS}$) is calculated via $\rm N_{H_2CS}$/$\rm N_{H_2}$. Apparently, $\rm f_{H_2CS}$ varies similarly to $\rm N_{H_2CS}$. Numerous studies have analyzed $\rm N_{H_2CS}$ and $\rm f_{H_2CS}$ in diverse massive star formation regions. Table~\ref{tab:CDA} lists the results of previous studies on H$_{2}$CS in line-rich cores of massive star formation regions. We combine these 7 line-rich cores with our 28 line-rich core samples, resulting in 35 line-rich core samples. Among the line-rich cores, Mon R 2 IRS 3 A \citep{2021MNRAS.507.1886F} and Sgr B2 \citep{2021Herschel} have lower $\rm N_{H_2CS}$ than the line-rich cores studied here. Except for these 2 line-rich cores, the other 33 have similar $\rm N_{H_2CS}$ ranges from $4.5\times10^{14}$ to $3.6\times10^{16}$ $\rm cm^{-2}$. These results indicate that $\rm N_{H_2CS}$ varies greatly among different line-rich cores within and outside the Galaxy.

Similarly, due to physical differences, $\rm N_{H_2}$ can vary greatly among different interstellar environments, leading to variation in $\rm f_{H_2CS}$. It is worth noting that I16351 core7 has a very high $\rm f_{H_2CS}$ of $1.6\times10^{-7}$. The reason for this large value is its high $\rm N_{H_2CS}$ and low $\rm N_{H_2}$. Except for this line-rich core, the abundance values of the other 34 are within the same range, from $1.7\times10^{-10}$ to $5.1\times10^{-8}$.

\subsection{Temperature-Abundance Relation}

Through the results collated in Table~\ref{tab:H2CSfit}, we can clearly find that temperature has a significant effect on the value of H$_{\rm 2}$CS abundance. Considering two variables X and Y, the formula for Pearson's correlation coefficient r can be expressed as:

\begin{equation}
 \label{eq:pearson}
 r = \frac{\sum^{n}_{i=1}(X_{i}-\bar{X})(Y_{i}-\bar{Y})}{(\sum^{n}_{i=1}(X_{i}-\bar{X})^{2})^{1/2}(\sum^{n}_{i=1}(Y_{i}-\bar{Y})^{2})^{1/2}},
\end{equation}
where $\rm \bar{X}$ and $\rm \bar{Y}$ are the mean values of X and Y. The value of Pearson's correlation coefficient for temperature and $\rm f_{H_2CS}$ is 0.71. The value of $\rm f_{H_2CS}$ relies on temperature, and the corresponding trend line shown in left panel of Figure \ref{X_fig} is fitted as,

\begin{equation}
 \label{eq:coeficient1}
  Log(f_{\text{$H_2CS$}}) = 0.02 \times T - 9.76,
\end{equation}
with uncertainties of 9.4\% and 1.3\% for slope and intercept. We find the abundance tends to increase with increasing temperature. H${_2}$CS molecule is thought to be trapped on icy dust grains and principally form on dust grain surface at low temperatures \citep{2020ARA&A..58..727J}, and is generally desorbed from the grain surface into gas-phase as the temperature increases \citep{2018MNRAS.474.5575V, 2019MNRAS.486.5197V}. The chemical property of H${_2}$CS in the hot regions is to a high degree inherited from the surrounding cold dust regions \citep{2022A&A...659A.100E}. Therefore $\rm f_{H_2CS}$ would be a good tracer of gas temperature.

\subsection{Column Density-Abundance Relation}

\begin{figure*}[ht!]
\centering
\includegraphics[width=0.48\linewidth]{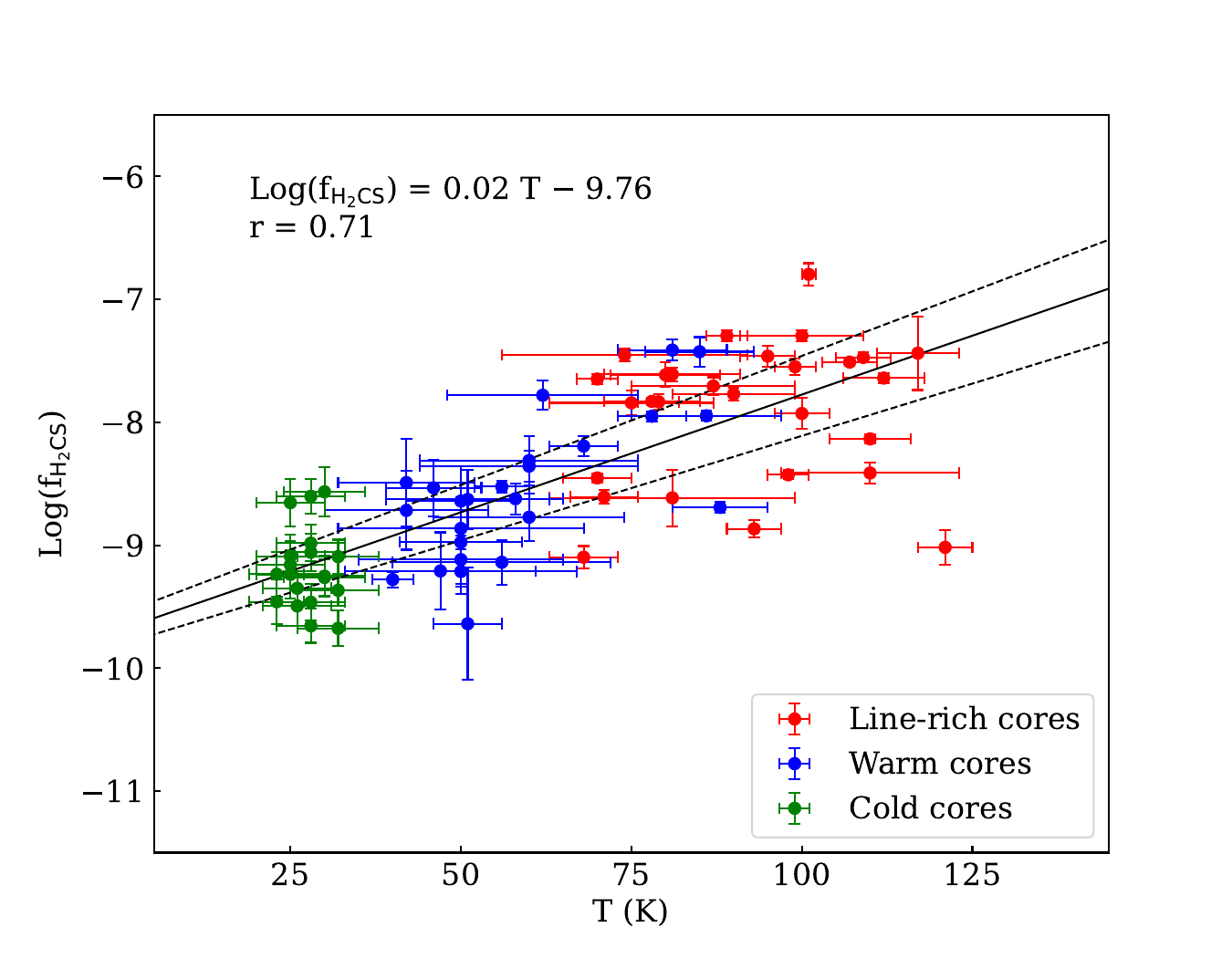}
\includegraphics[width=0.48\linewidth]{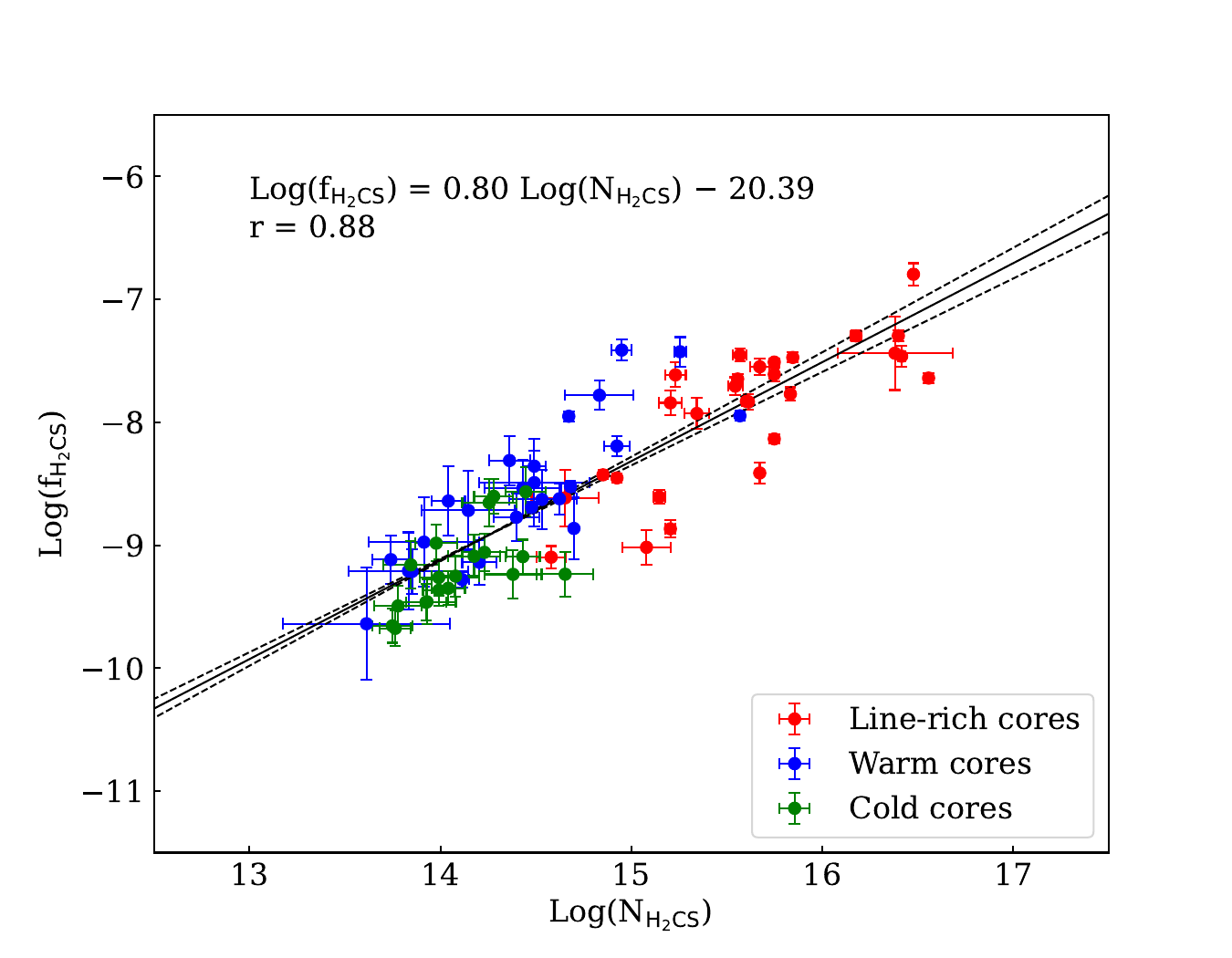}
\caption{The left panel shows the correlation between temperature and $\rm f_{H_2CS}$ and the right panel shows the correlation between $\rm N_{H_2CS}$ and $\rm f_{H_2CS}$, which are examined in this study. The data points are represented by different colored dots: red dots represent line-rich cores, blue dots represent warm cores and green dots represent cold cores. The best fitted line is shown by a black slash. The dashed lines are the fitted lines at maximum uncertainties.}
\label{X_fig}
\end{figure*}

Since $\rm N_{H_2CS}$ and $\rm f_{H_2CS}$ show the same trend in all types of cores, it would be pertinent to discuss the relationship of $\rm N_{H_2CS}$ and $\rm f_{H_2CS}$. The value of Pearson's correlation coefficient for $\rm N_{H_2CS}$ and $\rm f_{H_2CS}$ is 0.88, indicating a high correlation between H$_2$CS and H$_2$.

The correlation between $\rm N_{H_2CS}$ and $\rm f_{H_2CS}$ is shown in right panel of Figure \ref{X_fig}. We can see that $\rm N_{H_2CS}$ is sensitive to changes in $\rm f_{H_2CS}$. Thus, the trend line of $\rm N_{H_2CS}$ and $\rm f_{H_2CS}$ can be expressed using the following equation,

\begin{equation}
 \label{eq:coeficient2}
  Log(f_{\text{$H_2CS$}}) = 0.80 \times Log(N_{\text{$H_2CS$}}) - 20.39,
\end{equation}
with uncertainties of 5.3\% and 3.1\% for slope and intercepts. The red dots represent line-rich cores that have abundant organic molecules and high temperatures, column densities, and abundances of H$_2$CS. The line-rich cores have higher $\rm N_{H_2CS}$ and $\rm f_{H_2CS}$ than warm cores and cold cores. Additionally, warm cores are largely scattered above the line, while cold cores are below the line. This indicates that warm cores have higher $\rm f_{H_2CS}$ than cold cores at same $\rm N_{H_2CS}$.

This trend suggests that different regions have varying chemical characteristics and that the H$_2$CS abundances are enhanced from cold cores, warm cores to line-rich cores. Additionally, this proportional relationship can provide insights into the distribution and formation of H$_2$CS in interstellar space. Further observations and studies are still needed to better understand the physical and chemical properties of H${_2}$CS in space.

\section{Conclusions} \label{sec:cite}
\label{con}

We have presented the continuum and H$_2$CS line observations using the ALMA Band-7 survey toward 11 massive protocluster sources. The observations have detected a total of 145 continuum dense cores. Of these, 72 have H$_2$CS emission, including 28 line-rich cores, 25 warm cores, and 19 cold cores. The major results are summarized as follows:

(1) 72 cores with H$_2$CS in our sample have column densities exceed a threshold value for star formation, stating that H$_{\rm 2}$CS can be widely distributed in star-forming regions with different physical environments.

(2) Spatial distribution of H$_2$CS are revealed among the 11 sources. H$_{\rm 2}$CS emissions come from extended areas around dense rich-line cores and warm cores, and can also cover large areas of cold cores.

(3) H$_{\rm 2}$CS is found extensively distributed in protoclusters and $\rm f_{H_2CS}$ increases with temperature from cold cores, warm cores to line-rich cores, indicating that H$_{\rm 2}$CS is a good tracer for temperature in a variety of complex physical environments.

(4) $\rm N_{H_2CS}$ and $\rm f_{H_2CS}$ are found tightly correlated among different types of cores,  showing that the H$_2$CS abundances are enhanced from cold cores, warm cores to line-rich cores in star forming regions.

Our ALMA observations of the H$_2$CS line provide a substantial dataset of massive protoclusters with high angular resolution, which plays a crucial role in studying the formation mechanism and spatial distribution of H$_2$CS in interstellar space. Further observations and studies will definitely enhance our understanding of the H$_2$CS formation network and contribute to advancements in astrochemistry.

\begin{acknowledgments}

This paper makes use of the following ALMA data: ADS/JAO.ALMA\#2017.1.00545.S. ALMA is a partnership of ESO (representing its member states), NSF (USA), and NINS (Japan), together with NRC (Canada), MOST and ASIAA (Taiwan), and KASI (Republic of Korea), in cooperation with the Republic of Chile. The Joint ALMA Observatory is operated by ESO, AUI/NRAO, and NAOJ. This work has been supported by National Key R\&D Program of China (No.2022YFA1603101), and by NSFC through the grants No.12033005, No.12073061, No.12122307, and No.12103045. S.-L. Qin thanks the Xinjiang Uygur Autonomous Region of China for their support through the Tianchi Program. MYT acknowledges the support by NSFC through grants No.12203011, and Yunnan provincial Department of Science and Technology through grant No.202101BA070001-261. T. Zhang thanks the student's exchange program of the Collaborative Research Centre 956, funded by the Deutsche Forschungsgemeinschaft (DFG).

\end{acknowledgments}

\vspace{5mm}
\facility{ALMA}

\software{astropy \citep{2013A&A...558A..33A, 2018AJ....156..123A, 2022ApJ...935..167A}, CASA \citep{2007ASPC..376..127M, 2022PASP..134k4501C}, XCLASS \citep{2017A&A...598A...7M}, MAGIX \citep{2013A&A...549A..21M}.}

\appendix

\section{Cores identification and 2D Gaussian fitting}

The complete results of cores identification of the 11 IRAS sources are shown in Figure~\ref{cores_id0} and Figure~\ref{cores_id1}. The background color maps show the ALMA 345 GHz continuum flux in J2000.0 coordinates. The overlaid black contours of flux start from 8 rms and increase with power-law function. Each central location of identified cloud core is marked with yellow cross and the corresponding ID number is labeled beside the cross. The deconvolved parameters of each core was obtained from 2D Gaussian fitting tool in CASA and the whole parameters of the 145 cores are list in Table~\ref{f:spec1}. 

\section{The fitted H\texorpdfstring{$_2$}CCS spectral lines}

Figure~\ref{f:spec1} illustrates the 72 observed molecular lines and the synthetic spectra of the best fitting parameters for H$_2$CS lines. All the nine transitions of H$_2$CS surveyed in ALMA Band-7, which are summarized in Table~\ref{tab:H2CS}, are only tuned in SPW 31 and so here we present frequency range from roughly 342.95 GHz to 344.00 GHz.

\begin{figure*}
\newcounter{1}
\setcounter{1}{\value{figure}}
\setcounter{figure}{0}
\renewcommand\thefigure{A\arabic{figure}}
\centering
\includegraphics[width=\linewidth]{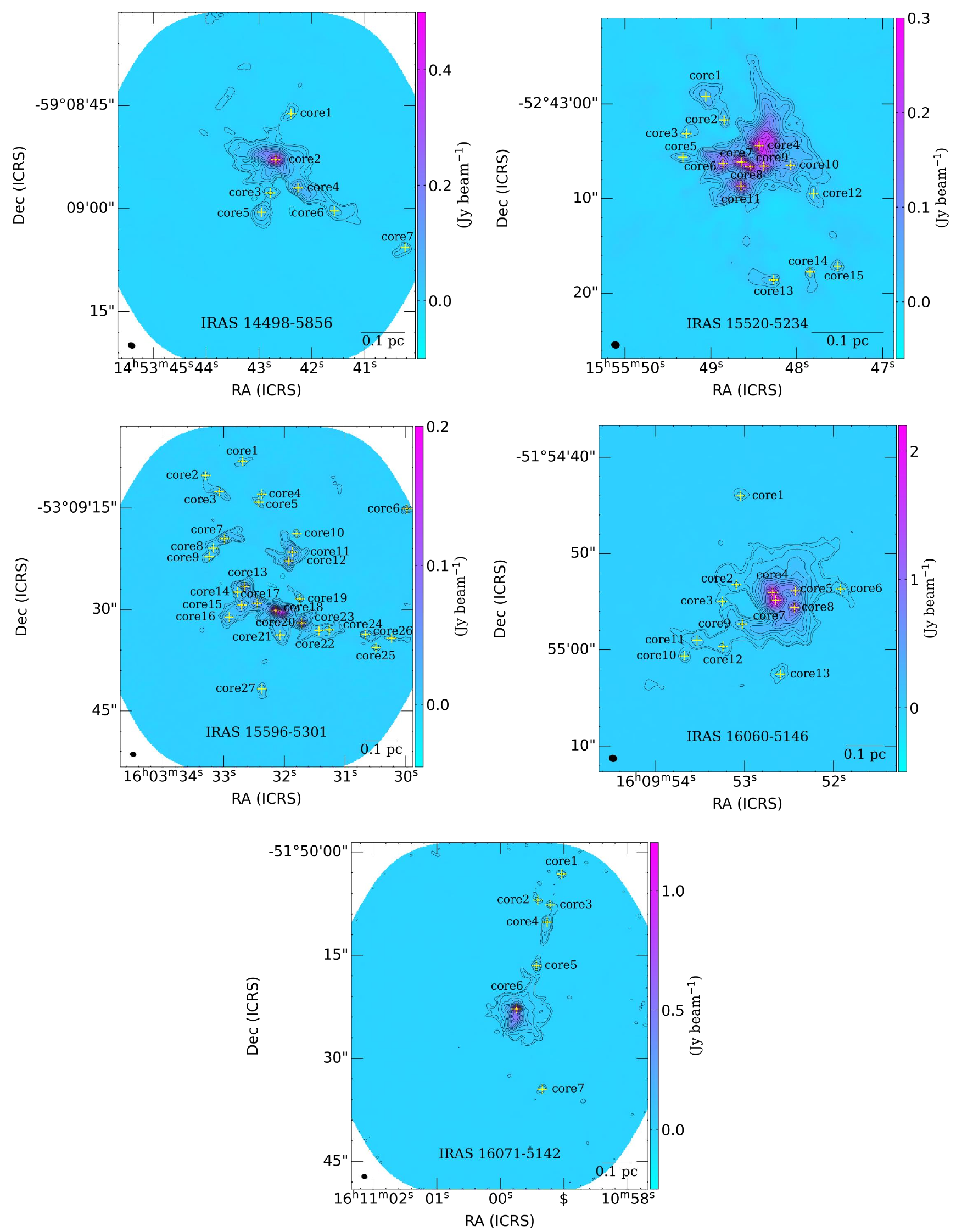}
\caption{The same as Figure \ref{cores_id0} but for the other researched 10 sources. The magnitude of the noise level of the continuum maps associated with each source is as follows: 1.0, 1.8, 0.5, 2.5, 1.3, 0.6, 1.5, 1.3, 0.6 and 1.7 mJy beam$^{-1}$.}
\label{cores_id1}
\setcounter{figure}{\value{1}}
\end{figure*}

\begin{figure*}
\setcounter{1}{\value{figure}}
\setcounter{figure}{0}
\renewcommand\thefigure{A\arabic{figure}}
\centering
\includegraphics[width=\linewidth]{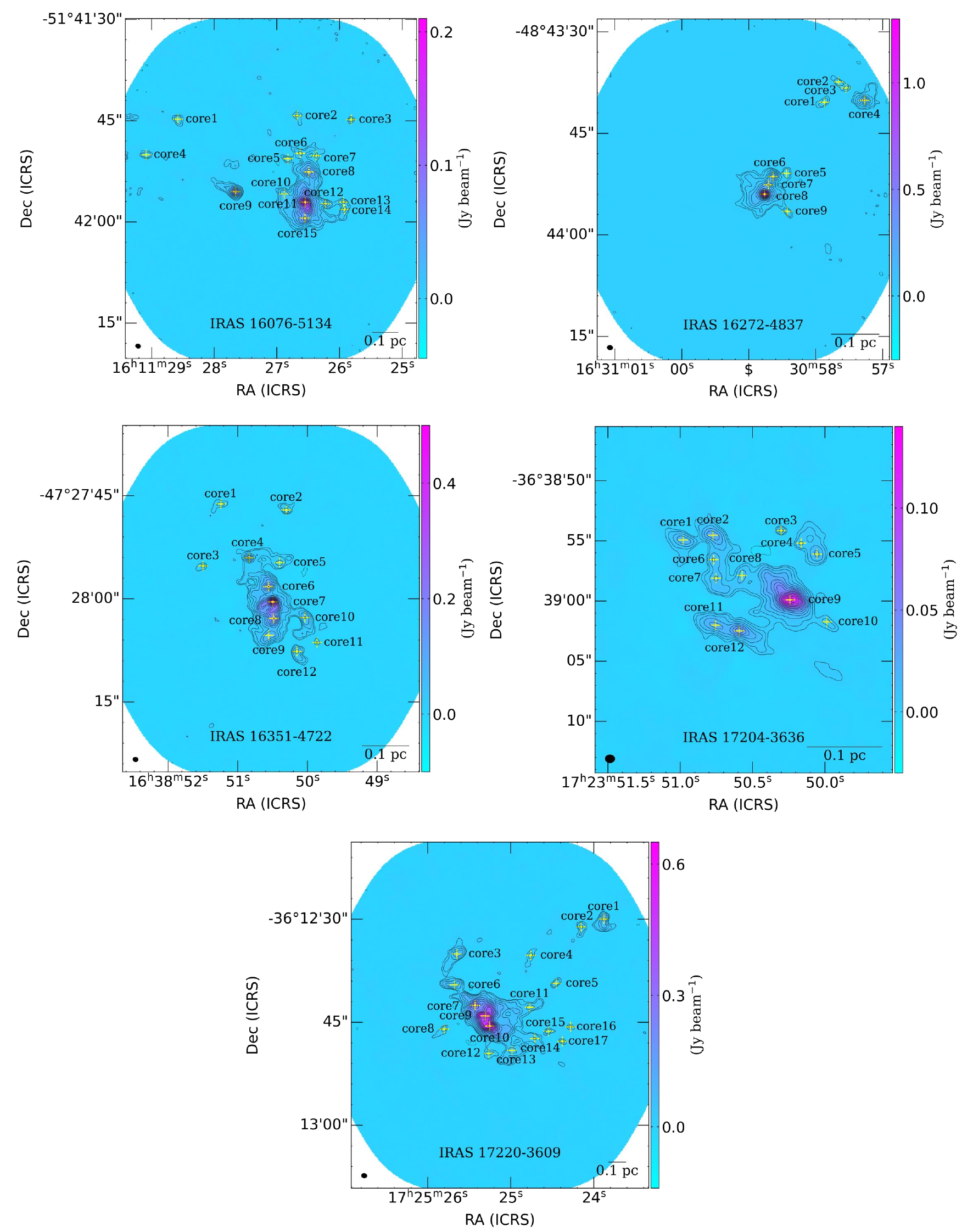}
\caption{(continued)}
\label{cores_id2}
\setcounter{figure}{\value{1}}
\end{figure*}

\clearpage

\begin{figure*}
\newcounter{2}
\setcounter{2}{\value{figure}}
\setcounter{figure}{1}
\renewcommand\thefigure{A\arabic{figure}}
\centering
\includegraphics[width=\linewidth]{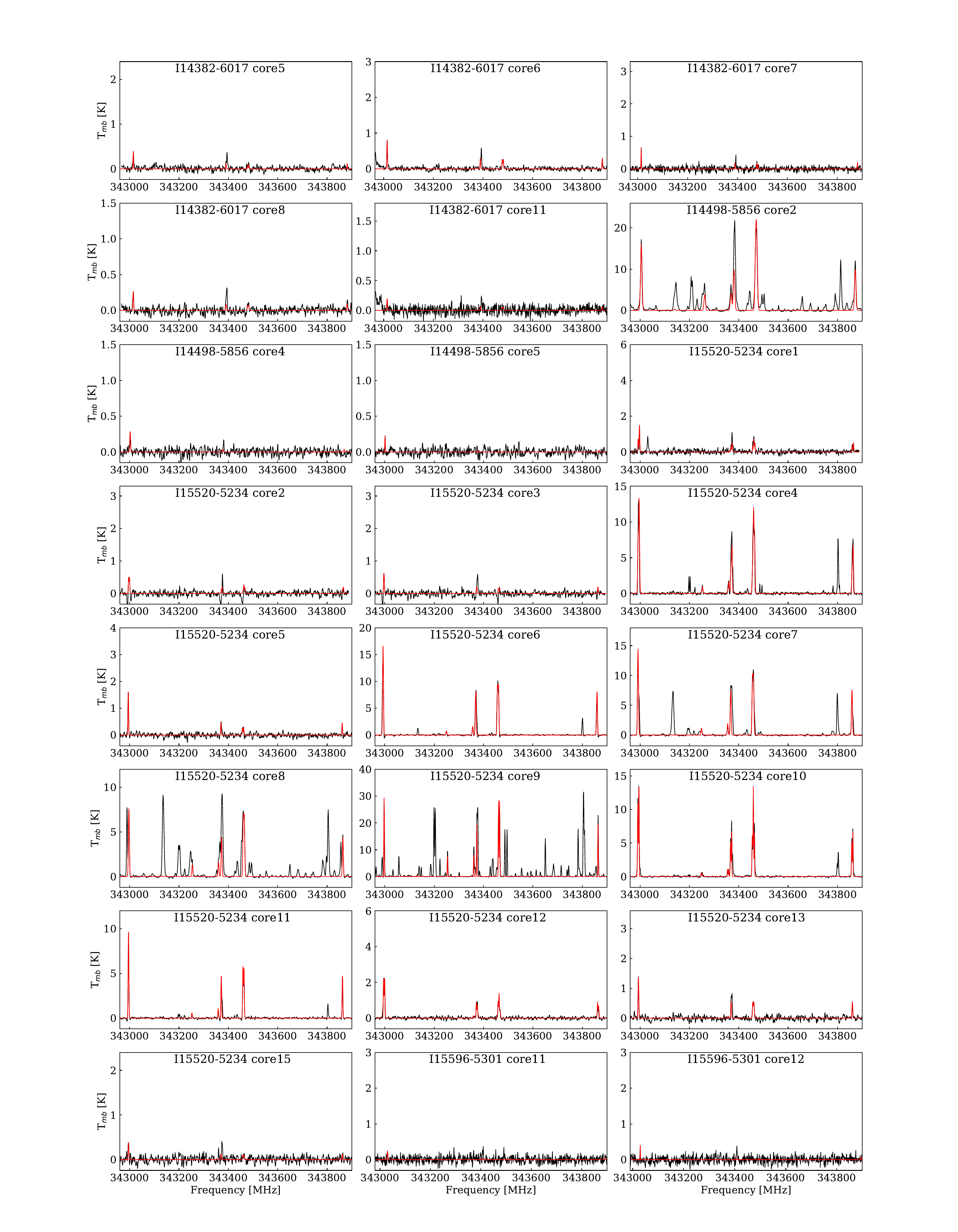}
\caption{The black lines are observed molecular lines and the red lines are synthetic spectra of the best fitting parameters for H$_{2}$CS lines.}
\label{f:spec1}
\setcounter{figure}{\value{2}}
\end{figure*}

\begin{figure*}
\setcounter{2}{\value{figure}}
\setcounter{figure}{1}
\renewcommand\thefigure{A\arabic{figure}}
\centering
\includegraphics[width=\linewidth]{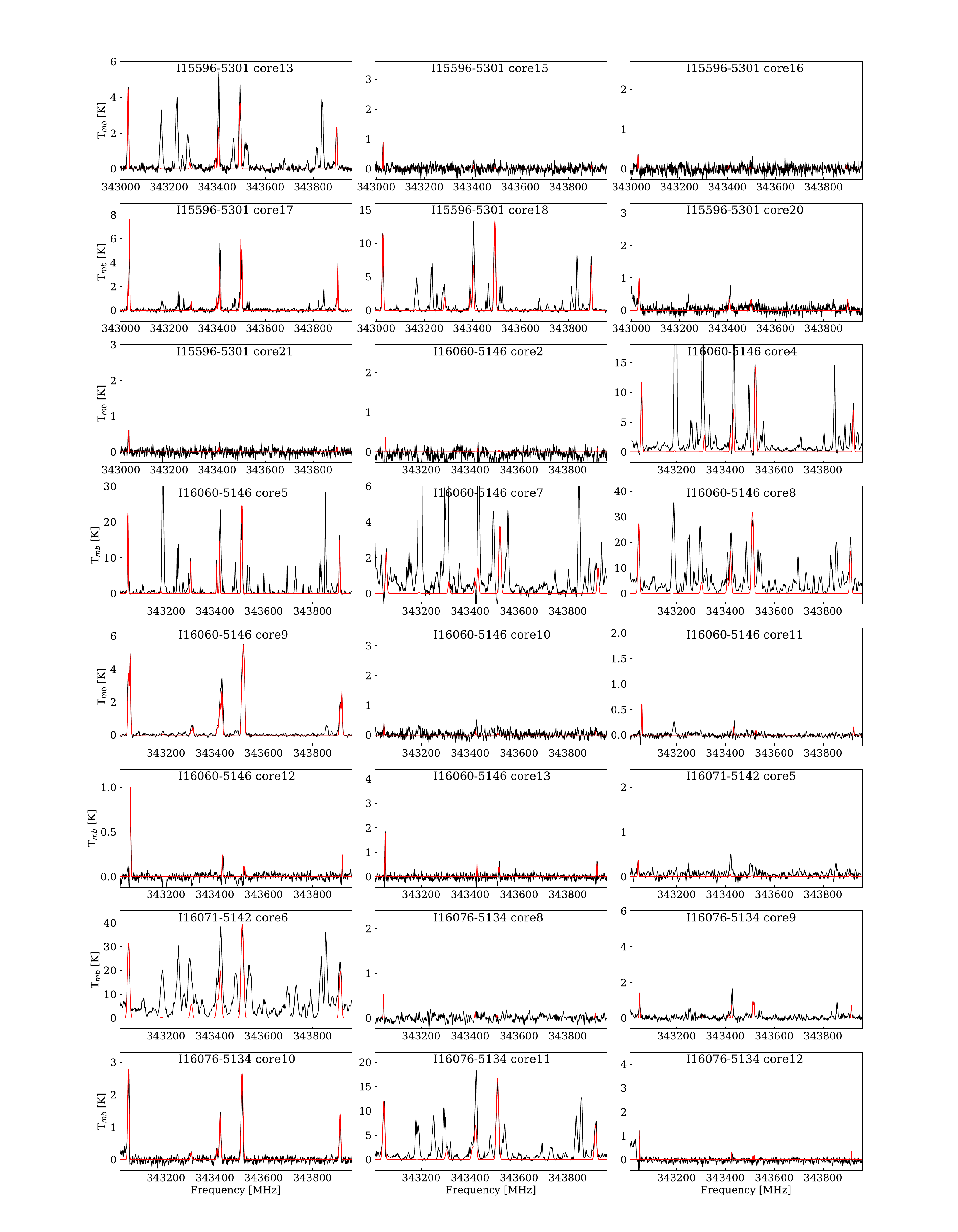}
\caption{(continued)}
\label{f:spec2}
\setcounter{figure}{\value{2}}
\end{figure*}

\begin{figure*}
\setcounter{2}{\value{figure}}
\setcounter{figure}{1}
\renewcommand\thefigure{A\arabic{figure}}
\centering
\includegraphics[width=\linewidth]{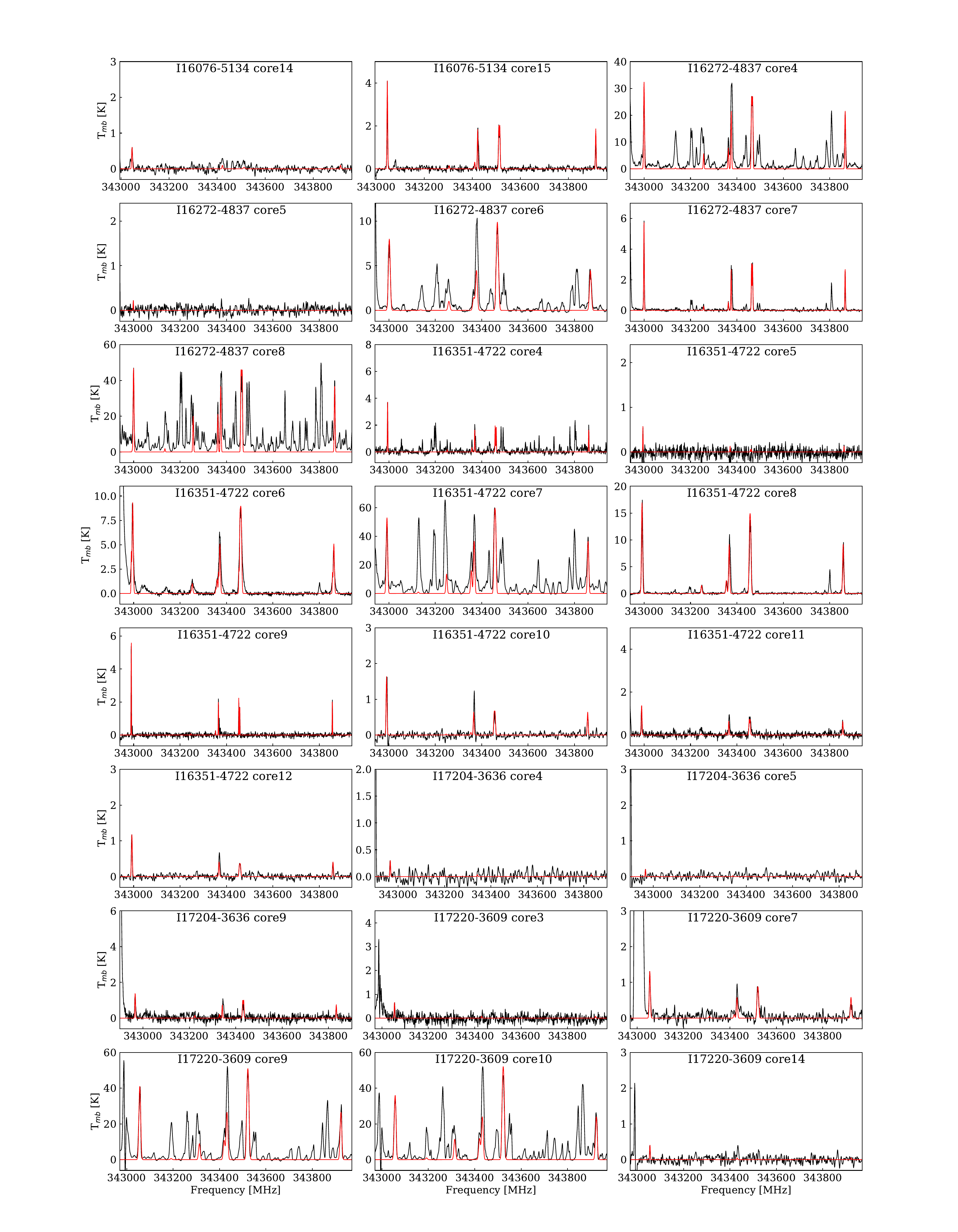}
\caption{(continued)}
\label{f:spec3}
\setcounter{figure}{\value{2}}
\end{figure*}

\clearpage

\startlongtable
\begin{deluxetable*}{ccccccccccc}
\tabletypesize{\scriptsize}
% \tabletypesize{\footnotesize}
\tablewidth{0pt} 
\tablenum{A1}
\tablecaption{2D Gaussian Fit Results}
\label{tab:Gaussfit}
\tablehead{
\colhead{ID} & \colhead{IRAS}& \colhead{Core} & \colhead{RA} & \colhead{DEC} & \colhead{Source size} & \colhead{PA} & \colhead{Mass} & \colhead{Integrated flux} & \colhead{Peak intensity} & Note\\
\colhead{} & \colhead{} & \colhead{} & \colhead{} & \colhead{} & \colhead{($\arcsec\times\arcsec$)} & \colhead{($^{\circ}$)} & \colhead{($\rm M_{\odot}$)} & \colhead{(mJy)} & \colhead{(mJy beam$^{-1}$)}
}
\colnumbers
\startdata
1 & I14382-6017 & 1 & 14:42:03.6 & $-$60:30:10.4 & 1.5$\times$1.3 & 51 & 25.4±5.0 & 127±11 & 42±3 \\
2 & I14382-6017 & 2 & 14:42:02.5 & $-$60:30:10.2 & 2.4$\times$1.1 & 100 & 26.6±5.0 & 133±7 & 33±1 \\
3 & I14382-6017 & 3 & 14:42:01.9 & $-$60:30:09.6 & 1.2$\times$0.6 & 68 & 4.6±1.1 & 23±4 & 13±1 \\
4 & I14382-6017 & 4 & 14:42:02.9 & $-$60:30:22.8 & 1.0$\times$0.6 & 90 & 7.8±1.4 & 39±2 & 24±1 \\
5 & I14382-6017 & 5 & 14:42:03.1 & $-$60:30:26.1 & 1.6$\times$0.8 & 69 & 1.4±0.5 & 13±3 & 52±1 \\
6 & I14382-6017 & 6 & 14:42:02.8 & $-$60:30:27.3 & 1.5$\times$1.0 & 33 & 32.4±9.6 & 380±30 & 142±9 \\
7 & I14382-6017 & 7 & 14:42:02.6 & $-$60:30:29.5 & 1.5$\times$1.2 & 43 & 20.5±7.0 & 210±8 & 71±2 \\
8 & I14382-6017 & 8 & 14:42:02.4 & $-$60:30:31.3 & 2.2$\times$1.0 & 10 & 15.6±4.7 & 160±6 & 42±1 \\
9 & I14382-6017 & 9 & 14:42:02.1 & $-$60:30:34.8 & 2.8$\times$1.3 & 65 & 41.8±8.8 & 209±23 & 44±4 \\
10 & I14382-6017 & 10 & 14:42:02.0 & $-$60:30:36.8 & 1.9$\times$0.9 & 18 & 14.6±2.6 & 73±2 & 23±1 \\
11 & I14382-6017 & 11 & 14:42:02.1 & $-$60:30:44.5 & 1.7$\times$1.1 & 174 & 32.6±4.0 & 343±26 & 104±6 \\
12 & I14498-5859 & 1 & 14:53:42:4 & $-$59:08:46.1 & 3.6$\times$1.3 & 144 & 4.1±0.8 & 108±8 & 15±1 \\
13 & I14498-5859 & 2 & 14:53:42.7 & $-$59:08:52:8 & 3.2$\times$1.9 & 65 & 21.4±1.7 &  2830±200 & 340±22 & * \\
14 & I14498-5859 & 3 & 14:53:42.8 & $-$59:08:57.7 & 1.6$\times$1.5 & 40 & 3.3±0.9 & 87±17 & 23±4 \\
15 & I14498-5859 & 4 & 14:53:42.2 & $-$59:08:57.2 & 4.1$\times$1.6 & 57 & 27.3±5.7 & 716±60 & 80±6 \\
16 & I14498-5859 & 5 & 14:53:42.9 & $-$59:09:00.6 & 1.9$\times$1.4 & 148 & 8.4±1.7 & 219±15 & 50±3 \\
17 & I14498-5859 & 6 & 14:53:41.6 & $-$59:09:00.2 & 3.0$\times$1.5 & 50 & 7.4±1.5 & 193±14 & 30±2 \\
18 & I14498-5859 & 7 & 14:53:40.3 & $-$59:09:06.0 & 2.8$\times$1.6 & 121 & 4.0±0.9 & 104±11 & 16±2 \\
19 & I15520-5234 & 1 & 15:55:49.1 & $-$52:42:59.2 & 2.9$\times$1.7 & 40 & 3.9±1.1 & 399±31 & 42±3 \\
20 & I15520-5234 & 2 & 15:55:48.8 & $-$52:43:01.7 & 1.4$\times$0.4 & 16 & 1.0±0.2 &  99±4 & 41±1 \\
21 & I15520-5234 & 3 & 15:55:49.3 & $-$52:43:02.9 & 2.0$\times$1.3 & 115 & 1.5±0.3 & 128±13 & 23±2 \\
22 & I15520-5234 & 4 & 15:55:48.4 & $-$52:43:04.2 & 3.7$\times$2.5 & 19 & 30.2±3.2 & 4130±230 & 250±13 & * \\
23 & I15520-5234 & 5 & 15:55:49.3 & $-$52:43:05.7 & 2.1$\times$0.7 & 87 & 1.7±0.4 & 115±7 & 32±2 \\
24 & I15520-5234 & 6 & 15:55:48.9 & $-$52:43:06.1 & 2.9$\times$2.0 & 158 & 11.2±1.0 & 1600±120 & 149±11 & * \\
25 & I15520-5234 & 7 & 15:55:48.7 & $-$52:43:06.1 & 1.9$\times$1.1 & 76 & 7.0±1.0 & 968±137 & 217±26 & * \\
26 & I15520-5234 & 8 & 15:55:48.5 & $-$52:43:06.8 & 1.6$\times$1.2 & 60 & 4.3±0.7 & 819±133 & 197±26 & * \\
27 & I15520-5234 & 9 & 15:55:48.4 & $-$52:43:06.5 & 2.0$\times$1.4 & 119 & 6.1±0.3 & 1128±35 & 200±5 & * \\
28 & I15520-5234 & 10 & 15:55:48.1 & $-$52:43:06.5 & 1.9$\times$1.4 & 85 & 3.4±0.8 & 440±18 & 80±3 & * \\
29 & I15520-5234 & 11 & 15:55:48.6 & $-$52:43:08.5 & 3.1$\times$1.9 & 63 & 7.9±1.6 & 1510±190 & 142±17 & * \\
30 & I15520-5234 & 12 & 15:55:47.8 & $-$52:43:09.7 & 2.8$\times$1.3 & 23 & 1.8±0.4 & 190±19 & 26±2 \\
31 & I15520-5234 & 13 & 15:55:48.3 & $-$52:43:18.6 & 3.4$\times$1.2 & 80 & 2.3±0.7 & 234±27 & 29±3 \\
32 & I15520-5234 & 14 & 15:55:47.8 & $-$52:43:18.3 & 2.6$\times$0.8 & 0 & 2.6±0.7 & 125±21 & 24±3 \\
33 & I15520-5234 & 15 & 15:55:47.5 & $-$52:43:17.1 & 1.2$\times$0.9 & 88 & 1.0±0.2 & 81±8 & 29±2           \\
34 & I15596-5301 & 1 & 16:03:32.7 & $-$53:09:08.2 & 2.0$\times$1.1 & 117 & 11.7±2.3 & 34±3 & 7±1 \\
35 & I15596-5301 & 2 & 16:03:33.3 & $-$53:09:10.2 & 1.2$\times$0.9 & 167 & 7.6±1.5 & 22±2 & 8±1 \\
36 & I15596-5301 & 3 & 16:03:33.1 & $-$53:09:12.6 & 1.9$\times$0.7 & 62 & 14.1±2.7 & 41±3 & 11±1 \\
37 & I15596-5301 & 4 & 16:03:32.4 & $-$53:09:13.0 & 1.7$\times$0.8 & 145 & 7.9±1.5 & 23±1 & 6±1 \\
38 & I15596-5301 & 5 & 16:03:32.4 & $-$53:09:14.1 & 1.6$\times$0.8 & 55 & 9.3±1.8 & 27±2 & 8±1 \\
39 & I15596-5301 & 6 & 16:03:29.9 & $-$53:09:15.2 & 0.7$\times$0.3 & 116 & 8.9±1.7 & 26±2 & 18±1 \\
40 & I15596-5301 & 7 & 16:03:32.9 & $-$53:09:19.5 & 1.6$\times$0.7 & 111 & 28.2±5.1 & 82±2 & 25±1 \\
41 & I15596-5301 & 8 & 16:03:33.2 & $-$53:09:21.1 & 1.9$\times$1.0 & 12 & 22.0±4.2 & 64±4 & 14±1 \\
42 & I15596-5301 & 9 & 16:03:33.2 & $-$53:09:22.2 & 2.0$\times$1.0 & 115 & 16.9±3.1 & 49±2 & 10±1 \\
43 & I15596-5301 & 10 & 16:03:31.8 & $-$53:09:18.8 & 1.0$\times$0.7 & 24 & 6.5±1.2 & 19±1 & 9±1 \\
44 & I15596-5301 & 11 & 16:03:32.6 & $-$53:09:26.7 & 1.5$\times$1.0 & 76 & 61.6±11.1 & 179±4 & 51±5 \\
45 & I15596-5301 & 12 & 16:03:31.9 & $-$53:09:22.8 & 1.4$\times$0.9 & 139 & 53.7±10.4 & 156±12 & 49±3 \\
46 & I15596-5301 & 13 & 16:03:32.6 & $-$53:09:26.6 & 1.3$\times$0.8 & 67 & 12.3±1.7 & 160±16 & 100±23 & * \\
47 & I15596-5301 & 14 & 16:03:32.8 & $-$53:09:27.5 & 1.6$\times$0.6 & 16 & 20.0±3.7 & 58±3 & 19±1 \\
48 & I15596-5301 & 15 & 16:03:32.7 & $-$53:09:29.3 & 1.5$\times$0.9 & 177 & 44.1±7.9 & 128±3 & 36±1 \\
49 & I15596-5301 & 16 & 16:03:32.9 & $-$53:09:31.1 & 1.8$\times$1.3 & 101 & 35.8±6.5 & 104±4 & 21±1 \\
50 & I15596-5301 & 17 & 16:03:32.4 & $-$53:09:29.1 & 2.3$\times$1.3 & 60 & 24.1±6.9 & 263±71 & 42±10 \\
51 & I15596-5301 & 18 & 16:03:32.1 & $-$53:09:30.3 & 2.2$\times$0.9 & 66 & 55.4±4.1 & 799±54 & 110±39 & * \\
52 & I15596-5301 & 19 & 16:03:31.7 & $-$53:09:28.3 & 0.7$\times$0.3 & 24 & 6.9±1.4 & 20±2 & 14±1 \\
53 & I15596-5301 & 20 & 16:03:31.7 & $-$53:09:32.1 & 0.8$\times$0.6 & 63 & 29.5±11.9 & 176±32 & 95±12 \\
54 & I15596-5301 & 21 & 16:03:32.1 & $-$53:09:33.7 & 2.0$\times$1.4 & 12 & 35.8±6.5 & 104±4 & 18±1 \\
55 & I15596-5301 & 22 & 16:03:31.4 & $-$53:09:33.1 & 1.4$\times$1.1 & 91 & 16.5±3.1 & 48±3 & 13±1 \\
56 & I15596-5301 & 23 & 16:03:31.3 & $-$53:09:32.9 & 1.9$\times$1.3 & 65 & 22.0±4.0 & 64±2 & 12±1 \\
57 & I15596-5301 & 24 & 16:03:30.7 & $-$53:09:33.8 & 1.2$\times$0.7 & 70 & 12.7±2.8 & 37±5 & 15±1 \\
58 & I15596-5301 & 25 & 16:03:30.5 & $-$53:09:35.6 & 1.0$\times$0.5 & 75 & 5.9±1.1 & 17±1 & 9±1 \\
59 & I15596-5301 & 26 & 16:03:30.2 & $-$53:09:34.3 & 2.1$\times$0.4 & 79 & 7.9±1.5 & 23±1 & 7±1 \\
60 & I15596-5301 & 27 & 16:03:32.4 & $-$53:09:41.9 & 2.3$\times$1.0 & 7 & 16.2±3.1 & 47±3 & 8±1 \\
61 & I16060-5146 & 1 & 16:09:53.1 & $-$51:54:43.7 & 2.3$\times$1.7 & 114 & 17.8±3.9 & 223±26 & 28±3 \\
62 & I16060-5146 & 2 & 16:09:53.1 & $-$51:54:53.2 & 1.6$\times$1.1 & 119 & 22.5±4.4 & 282±16 & 66±3 \\
63 & I16060-5146 & 3 & 16:09:53.2 & $-$51:54:55.0 & 1.5$\times$0.7 & 72 & 8.4±1.6 & 105±5 & 36±1 \\
64 & I16060-5146 & 4 & 16:09:52.7 & $-$51:54:53.9 & 1.6$\times$1.1 & 11 & 99.2±17.0 & 6520±810 & 1530±160 & * \\
65 & I16060-5146 & 5 & 16:09:52.4 & $-$51:54:53.8 & 2.2$\times$1.2 & 34 & 93.6±8.0 & 5000±330 & 843±48 & * \\
66 & I16060-5146 & 6 & 16:09:51.9 & $-$51:54:53.7 & 1.5$\times$0.7 & 58 & 16.7±3.3 & 210±11 & 70±3 \\
67 & I16060-5146 & 7 & 16:09:52.6 & $-$51:54:55.2 & 1.6$\times$1.3 & 177 & 120.5±15.7 & 6990±880 & 1460±150 & * \\
68 & I16060-5146 & 8 & 16:09:52.4 & $-$51:54:55.6 & 1.6$\times$1.2 & 142 & 105.0±11.4 & 5500±550 & 1220±100 & * \\
69 & I16060-5146 & 9 & 16:09:53.0 & $-$51:54:57.4 & 1.4$\times$0.6 & 83 & 12.7±1.6 & 507±8 & 188±2 \\
70 & I16060-5146 & 10 & 16:09:53.7 & $-$51:55:00.7 & 1.1$\times$0.8 & 13 & 9.3±1.8 & 117±6 & 45±2 \\
71 & I16060-5146 & 11 & 16:09:53.5 & $-$51:54:59.0 & 1.7$\times$1.2 & 80 & 23.4±2.3 & 389±25 & 85±5 \\
72 & I16060-5146 & 12 & 16:09:53.2 & $-$51:54:59.6 & 1.5$\times$0.8 & 58 & 18.6±3.6 & 234±11 & 68±3 \\
73 & I16060-5146 & 13 & 16:09:52.6 & $-$51:55:02.4 & 2.0$\times$1.2 & 137 & 10.3±1.7 & 203±12 & 37±2 \\
74 & I16071-5142 & 1 & 16:10:59.0 & $-$51:50:03.2 & 0.9$\times$0.7 & 45 & 7.7±1.7 & 62±8 & 29±3 \\
75 & I16071-5142 & 2 & 16:10:59.4 & $-$51:50:06.9 & 1.5$\times$0.4 & 50 & 5.9±1.2 & 48±5 & 20±1 \\
76 & I16071-5142 & 3 & 16:10:59.2 & $-$51:50:07.7 & 0.9$\times$0.7 & 54 & 5.9±1.1 & 48±2 & 23±1 \\
77 & I16071-5142 & 4 & 16:10:59.3 & $-$51:50:10.5 & 2.1$\times$0.6 & 180 & 28.7±5.5 & 232±18 & 56±4 \\
78 & I16071-5142 & 5 & 16:10:59.4 & $-$51:50:16.6 & 0.7$\times$0.5 & 0 & 12.6±2.3 & 102±6 & 60±2 \\
79 & I16071-5142 & 6 & 16:10:59.8 & $-$51:50:23.1 & 1.8$\times$1.0 & 164 & 62.9±9.9 & 3840±580 & 849±107 & * \\
80 & I16071-5142 & 7 & 16:10:59.4 & $-$51:50:34.5 & 1.5$\times$0.6 & 139 & 8.9±1.8 & 72±7 & 24±2 \\
81 & I16076-5134 & 1 & 16:11:28.6 & $-$51:41:44.9 & 2.0$\times$1.4 & 57 & 5.4±1.3 & 62±9 & 9±1 \\
82 & I16076-5134 & 2 & 16:11:26.7 & $-$51:41:44.6 & 3.2$\times$1.0 & 33 & 5.0±1.3 & 58±9 & 7±1 \\
83 & I16076-5134 & 3 & 16:11:25.8 & $-$51:41:44.9 & 1.2$\times$0.6 & 149 & 2.1±0.5 & 24±4 & 8±1 \\
84 & I16076-5134 & 4 & 16:11:29.1 & $-$51:41:50.0 & 1.5$\times$0.9 & 86 & 3.5±0.8 & 40±5 & 10±1 \\
85 & I16076-5134 & 5 & 16:11:26.8 & $-$51:41:50.7 & 2.0$\times$0.8 & 69 &  0.9±0.3 & 10±2 & 2±1 \\
86 & I16076-5134 & 6 & 16:11:26.6 & $-$51:41:50.2 & 2.9$\times$1.9 & 39 & 12.7±3.0 & 147±19 & 11±1 \\
87 & I16076-5134 & 7 & 16:11:26.4 & $-$51:41:50.2 & 2.2$\times$1.2 & 161 & 4.8±1.0 & 55±1 & 8±1 \\
88 & I16076-5134 & 8 & 16:11:26.5 & $-$51:41:52.7 & 2.3$\times$1.7 & 71 & 38.9±8.3 & 449±32 & 49±3 \\
89 & I16076-5134 & 9 & 16:11:27.7 & $-$51:41:55.6 & 0.9$\times$0.7 & 69 & 5.4±1.2 & 218±5 & 93±1 & * \\
90 & I16076-5134 & 10 & 16:11:26.9 & $-$51:41:56.7 & 3.1$\times$1.7 & 18 & 5.7±0.8 & 214±21 & 17±2 \\
91 & I16076-5134 & 11 & 16:11:26.5 & $-$51:41:57.4 & 2.3$\times$1.5 & 16 & 30.5±2.6 & 1410±110 & 160±11 & * \\
92 & I16076-5134 & 12 & 16:11:26.2 & $-$51:41:57.3 & 1.8$\times$0.8 & 70 & 4.9±1.5 & 86±10 & 20±2 \\
93 & I16076-5134 & 13 & 16:11:25.9 & $-$51:41:57.1 & 1.6$\times$0.6 & 55 & 3.8±0.9 & 44±6 & 13±1 \\
94 & I16076-5134 & 14 & 16:11:25.9 & $-$51:41:58.3 & 1.4$\times$0.6 & 151 & 4.0±1.1 & 46±9 & 15±1 \\
95 & I16076-5134 & 15 & 16:11:26.6 & $-$51:41:59.5 & 1.5$\times$1.1 & 113 & 10.0±0.8 & 310±11 & 67±2 \\
96 & I16272-4837 & 1 & 16:30:57.9 & $-$48:43:40.4 & 2.6$\times$0.9 & 125 & 5.2±1.0 & 139±8 & 23±1 \\
97 & I16272-4837 & 2 & 16:30:57.7 & $-$48:43:37.5 & 1.8$\times$0.6 & 51 & 3.5±0.7 & 95±7 & 28±2 \\
98 & I16272-4837 & 3 & 16:30:57.6 & $-$48:43:38.3 & 1.4$\times$0.8 & 110 & 2.5±0.5 & 67±4 & 21±1 \\
99 & I16272-4837 & 4 & 16:30:57.3 & $-$48:43:40.1 & 1.0$\times$0.8 & 77 & 3.3±0.2 & 549±27 & 212±8 & * \\
100 & I16272-4837 & 5 & 16:30:58.4 & $-$48:43:50.9 & 1.5$\times$0.9 & 173 & 4.6±0.9& 123±10 & 32±2 \\
101 & I16272-4837 & 6 & 16:30:58.6 & $-$48:43:51.4 & 1.2$\times$0.6 & 146 & 4.0±0.5 & 638±49 & 252±14 & * \\
102 & I16272-4837 & 7 & 16:30:58.7 & $-$48:43:52.6 & 1.1$\times$0.6 & 128 & 5.2±0.4 & 658±19 & 280±6 & * \\
103 & I16272-4837 & 8 & 16:30:58.8 & $-$48:43:54.0 & 0.8$\times$0.8 & 161 & 14.0±1.2 & 2560±160 & 1144±51 & * \\
104 & I16272-4837 & 9 & 16:30:58.4 & $-$48:43:56.7 & 1.8$\times$0.5 & 35 & 3.3±0.6 & 88±4 & 27±1 \\
105 & I16351-4722 & 1 & 16:38:51.2 & $-$47:27:46.2 & 2.4$\times$1.1 & 112 & 3.1±0.7 & 111±12 & 18±2 \\
106 & I16351-4722 & 2 & 16:38:50.3 & $-$47:27:47.1 & 1.1$\times$0.8 & 167 & 2.6±0.6 & 95±10 & 34±3 \\
107 & I16351-4722 & 3 & 16:38:51.5 & $-$47:27:55.2 & 1.4$\times$0.9 & 62 & 1.8±0.4 & 65±4 & 18±4 \\
108 & I16351-4722 & 4 & 16:38:50.8 & $-$47:27:54.1 & 0.8$\times$0.6 & 52 & 1.7±0.1 & 220±9 & 113±3 & * \\
109 & I16351-4722 & 5 & 16:38:50.4 & $-$47:27:54.7 & 1.5$\times$1.0 & 118 & 4.0±0.8 & 146±4 & 35±1 \\
110 & I16351-4722 & 6 & 16:38:50.6 & $-$47:27:58.1 & 1.9$\times$1.6 & 8 & 8.1±1.3 & 1024±79 & 144±10 & * \\
111 & I16351-4722 & 7 & 16:38:50.5 & $-$47:28:00.8 & 2.2$\times$1.9 & 180 & 11.6±1.8 & 3010±470 & 320±45 & * \\
112 & I16351-4722 & 8 & 16:38:50.5 & $-$47:28:02.8 & 2.1$\times$1.6 & 159 & 11.4±1.4 & 1511±36 & 199±4 & * \\
113 & I16351-4722 & 9 & 16:38:50.6 & $-$47:28:05.3 & 2.7$\times$1.2 & 119 & 7.7±0.6 & 597±10 & 79±1 \\
114 & I16351-4722 & 10 & 16:38:50.0 & $-$47:28:02.8 & 2.0$\times$0.9 & 21 & 4.7±0.6 & 378±23 & 63±4 \\
115 & I16351-4722 & 11 & 16:38:49.9 & $-$47:28:06.2 & 0.8$\times$0.6 & 22 & 0.3±0.1 & 33±1 & 17±1 \\
116 & I16351-4722 & 12 & 16:38:50.1 & $-$47:28:07.7 & 1.3$\times$0.9 & 174 & 2.5±0.6 & 170±9 & 49±4 \\
117 & I17204-3636 & 1 & 17:23:51.0 & $-$36:38:55.0 & 1.7$\times$0.9 & 71 & 3.4±0.8 & 80±8 & 20±2 \\
118 & I17204-3636 & 2 & 17:23:50.8 & $-$36:38:54.6 & 2.1$\times$1.0 & 45 & 5.8±1.3 & 136±12 & 25±2 \\
119 & I17204-3636 & 3 & 17:23:50.3 & $-$36:38:54.1 & 0.8$\times$0.7 & 88 & 0.5±0.1 & 11±2 & 13±1 \\
120 & I17204-3636 & 4 & 17:23:50.2 & $-$36:38:55.1 & 1.2$\times$0.8 & 6 & 1.4±0.3 & 32±4 & 11±1 \\
121 & I17204-3636 & 5 & 17:23:50.1 & $-$36:38:56.0 & 1.5$\times$1.1 & 8 & 3.0±0.7 & 70±6 & 17±1 \\
122 & I17204-3636 & 6 & 17:23:50.8 & $-$36:38:56.6 & 1.3$\times$0.6 & 6 & 2.0±0.4 & 46±2 & 17±1 \\
123 & I17204-3636 & 7 & 17:23:50.7 & $-$36:38:58.4 & 1.7$\times$1.1 & 20 & 3.4±0.8 & 79±10 & 17±2 \\
124 & I17204-3636 & 8 & 17:23:50.6 & $-$36:38:58.0 & 2.2$\times$0.9 & 164 & 2.9±0.6 & 67±4 & 13±1 \\
125 & I17204-3636 & 9 & 17:23:50.2 & $-$36:38:59.9 & 2.0$\times$1.2 & 74 & 6.4±0.6 & 680±38 & 123±6 \\
126 & I17204-3636 & 10 & 17:23:50.0 & $-$36:39:01.9 & 2.4$\times$0.9 & 43 & 1.7±0.5 & 39±7 & 7±1 \\
127 & I17204-3636 & 11 & 17:23:50.8 & $-$36:39:01.9 & 2.0$\times$1.0 & 73 & 5.0±1.0 & 117±4 & 24±1 \\
128 & I17204-3636 & 12 & 17:23:50.6 & $-$36:39:02.5 & 2.3$\times$0.9 & 75 & 7.7±1.7 & 181±16 & 36±3 \\
129 & I17220-3609 & 1 & 17:25:23.9 & $-$36:12:30.4 & 2.0$\times$0.9 & 176 & 102.4±23.5 & 407±46 & 85±8 \\
130 & I17220-3609 & 2 & 17:25:24.1 & $-$36:12:31.2 & 1.9$\times$1.0 & 165 & 40.0±9.6 & 159±21 & 33±4 \\
131 & I17220-3609 & 3 & 17:25:25.6 & $-$36:12:35.0 & 2.0$\times$1.1 & 134 & 96.3±20.2 & 383±24 & 75±4 \\
132 & I17220-3609 & 4 & 17:25:24.8 & $-$36:12:35.6 & 3.7$\times$0.8 & 158 & 41.3±10.2 & 164±24 & 21±3 \\
133 & I17220-3609 & 5 & 17:25:24.5 & $-$36:12:39.3 & 1.6$\times$0.4 & 158 & 21.9±4.6 & 87±5 & 32±1 \\
134 & I17220-3609 & 6 & 17:25:25.7 & $-$36:12:39.5 & 2.7$\times$1.0 & 85 & 112.4±25.6 & 447±49 & 69±7 \\
135 & I17220-3609 & 7 & 17:25:25.4 & $-$36:12:42.6 & 1.8$\times$1.6 & 105 & 144.7±13.8 & 1960±120 & 295±16 & * \\
136 & I17220-3609 & 8 & 17:25:25.8 & $-$36:12:46.1 & 1.7$\times$0.6 & 135 & 21.6±4.8 & 86±8 & 25±2 \\
137 & I17220-3609 & 9 & 17:25:25.3 & $-$36:12:44.1 & 3.6$\times$1.4 & 83 & 263.1±25.7 & 5870±220 & 554±19 & * \\
138 & I17220-3609 & 10 & 17:25:25.3 & $-$36:12:45.4 & 2.6$\times$1.3 & 84 & 235.2±16.6 & 4560±220 & 595±26 & * \\
139 & I17220-3609 & 11 & 17:25:24.8 & $-$36:12:42.8 & 2.4$\times$0.8 & 111 & 54.1±11.1 & 215±9 & 43±2 \\
140 & I17220-3609 & 12 & 17:25:25.2 & $-$36:12:49.6 & 1.7$\times$1.2 & 96 & 30.2±9.1 & 120±27 & 24±5 \\
141 & I17220-3609 & 13 & 17:25:25.0 & $-$36:12:49.1 & 2.4$\times$1.5 & 89 & 82.8±17.7 & 329±25 & 43±3 \\
142 & I17220-3609 & 14 & 17:25:24.7 & $-$36:12:47.3 & 2.8$\times$1.3 & 62 & 70.9±14.4 & 282±9 & 36±1 \\
143 & I17220-3609 & 15 & 17:25:24.5 & $-$36:12:46.4 & 1.3$\times$0.5 & 123 & 16.6±3.8 & 66±7 & 28±2 \\
144 & I17220-3609 & 16 & 17:25:24.3 & $-$36:12:45.8 & 1.2$\times$0.7 & 79 & 13.8±3.0 & 55±5 & 22±1 \\
145 & I17220-3609 & 17 & 17:25:24.4 & $-$36:12:47.8 & 1.4$\times$0.6 & 46 & 15.6±3.7 & 62±8 & 23±2 \\
\enddata
\tablecomments{The rows marked with * indicate that the molecular cloud cores are recognized as line-rich cores.}
\end{deluxetable*}

\begin{deluxetable*}{cccccc}
\centering
\tablewidth{0pt} 
\tablenum{A2}
\tablecaption{H$_{2}$CS transitions in ALMA Band-7 survey}\label{tab:H2CS}
\tabletypesize{\footnotesize}
\tablehead{
\colhead{Frequency} & \colhead{Uncertainty} & \colhead{$J'_{K'_a,K'_c} - J'_{K'_a,K'_c}$} & \colhead{$\rm S_{\rm ij}$$\rm \mu^2$} & \colhead{$\rm Log_{\rm 10}(A_{\rm ij})$} & \colhead{E$_{\rm U}$} \\
\colhead{(MHz)} & \colhead{(MHz)} & \colhead{} & \colhead{(D$^{\rm 2}$)} & \colhead{($\rm s^{-1}$)} & \colhead{(K)}
}
\startdata
342946.4239 & 0.0500 & $\rm 10_{0,10}-9_{0,9}$ & 27.19603 & $-$3.21610 & 90.59115 \\
343203.2392 & 0.0500 & $\rm 10_{5,6}-9_{5,5}$ & 61.19507 & $-$3.34004 & 419.17248 \\
343203.2392 & 0.0500 & $\rm 10_{5,5}-9_{5,4}$ & 61.19507 & $-$3.34004 & 419.17248 \\
343309.8296 & 0.0500 & $\rm 10_{4,7}-9_{4,6}$ & 22.84414 & $-$3.29045 & 301.07181 \\
343309.8296 & 0.0500 & $\rm 10_{4,6}-9_{4,5}$  & 22.84414 & $-$3.29045 & 301.07181 \\
343322.0819 & 0.0500 & $\rm 10_{2,9}-9_{2,8}$ & 26.10686 & $-$3.23243 & 143.30653 \\
343409.9625 & 0.0500 & $\rm 10_{3,8}-9_{3,7}$ & 74.24497 & $-$3.25530 & 209.09441 \\
343414.1463 & 0.0500 & $\rm 10_{3,7}-9_{3,6}$ & 74.24322 & $-$3.25530 & 209.09476 \\
343813.1683 & 0.0500 & $\rm 10_{2,8}-9_{2,7}$ & 26.10949 & $-$3.23052 & 143.37729 \\
\enddata
\tablecomments{Data source: CDMS. The rest frequencies are listed with an uncertainty of 0.05 MHz. The $\rm 10_{0,10}-9_{0,9}$, $\rm 10_{3,8}-9_{3,7}$, $\rm 10_{3,7}-9_{3,6}$ and $\rm 10_{2,8}-9_{2,7}$ lines are not affected by other molecular lines and can be used for precise parameter estimation. In cold cores, only $\rm 10_{0,10}-9_{0,9}$ lines are detected. The $\rm 10_{5,6}-9_{5,5}$ and $\rm 10_{5,5}-9_{5,4}$ lines are blended with C$_2$H$_5$CN and NH$_2$CHO molecular lines in most line-rich cores but are not blended in warm cores. The $\rm 10_{4,7}-9_{4,6}$ and $\rm 10_{4,6}-9_{4,5}$ lines are blended with unidentified molecular lines in all line-rich cores and a few warm cores, and may also be blended with (CH$_2$OH)$_2$ molecular lines in some line-rich cores. The $\rm 10_{2,9}-9_{2,8}$ lines are blended with $\rm H_{2}^{13}CO$ in both line-rich cores and warm cores.}
\end{deluxetable*}

\begin{deluxetable*}{cccccc}
\centering
\tablewidth{0pt}
\tablenum{A3}
\tablecaption{The column densities and abundances of H$_{2}$CS}\label{tab:CDA}
\tabletypesize{\footnotesize}
\tablehead{
\colhead{Sources} & \colhead{Cores} & \colhead{$N_{H_2CS}$} & \colhead{$f_{H_2CS}$} & \colhead{Ref.} \\
\colhead{} & \colhead{} & \colhead{($cm^{\rm -2}$)} & \colhead{} & \colhead{}
}
\startdata
%LMC & N 105-1 A & \textless3.9$\times10^{13}$ & \textless1.7$\times10^{-10}$ & \cite{2022ApJ...931..102S} \\
%LMC & N 105-2 A & 2.8$\times10^{14}$ & 1.6$\times10^{-9}$ & \cite{2022ApJ...931..102S} \\
Mon R 2 & IRS 3 A & 8.0$\times10^{13}$  & 3.1$\times10^{-10}$ & \cite{2021MNRAS.507.1886F} \\
Sgr B2 & M & 2.5$\times10^{14}$ & 2.3$\times10^{-10}$ & \cite{2021Herschel} \\
Orion KL & MM1 & 8.0$\times10^{15}$ & 4.0$\times10^{-9}$ & \cite{2019ApJ...885...82L} \\
G33.92+0.11 & A5(1) & 7.2$\times10^{14}$ & 1.8$\times10^{-8}$ & \cite{2018ApJ...864..102M} \\
%LMC & ST11 & 2.8$\times10^{13}$ & 6.2$\times10^{-11}$ & \cite{2016ApJ...827...72S} \\
OMC 2 & FIR 4 & 6.7$\times10^{14}$ & 6.7$\times10^{-9}$ & \cite{2015ApJS..221...31S} \\
G9.62+0.19 & F & 2.6$\times10^{15}$ & 1.2$\times10^{-9}$ & \cite{2011ApJ...730..102L} \\
DR21(OH) & MM1a & 3.3$\times10^{15}$ & 4.0$\times10^{-9}$ & \cite{2011ApJ...737L..25M} \\
\enddata
\tablecomments{The column densities and abundances of H$_{2}$CS towards other 7 line-rich cores in 7 different massive star formation regions. The samples cover a variety of massive star-forming regions in the Galaxy, and the rich sample types facilitate our comparison. The shown $\rm N_{H_2CS}$ and $\rm f_{H_2CS}$ values are calculated from the peak center of all targets.}
\end{deluxetable*}

\clearpage

\bibliography{H2CS}{}
\bibliographystyle{aasjournal}

\end{document}